\documentclass[onecolumn,superscriptaddress,nobibnotes,aps,prd,nofootinbib,notitlepage,longbibliography]{revtex4-1}
\RequirePackage[english]{babel}
\RequirePackage[latin1]{inputenc}
\RequirePackage[T1]{fontenc}
\usepackage{graphicx,epsfig}
\usepackage{amsmath}
\usepackage {amssymb}
\usepackage {longtable}
\usepackage{multirow}

\usepackage{subfigmat}
\usepackage{BOONDOX-cal}

\usepackage{dsfont}
\usepackage{dcolumn}
\usepackage{bm}
\usepackage{amsfonts}
\usepackage{subfigure}
\usepackage{color}
\def\be{\begin{equation}}
\def\ee{\end{equation}}
\def\beq{\begin{eqnarray}}
\def\eeq{\end{eqnarray}}

\begin{document}
\title{Cosmological phase space of generalized hybrid metric-Palatini theories of gravity}

\author{Jo\~{a}o L. Rosa}
\email{joaoluis92@gmail.com}
\affiliation{Centro de Astrof\'{\i}sica e Gravita\c c\~ao - CENTRA,
Departamento de F\'{\i}sica,
Instituto Superior T\'{e}cnico - IST,
Universidade de Lisboa - UL,
Avenida Rovisco Pais 1, 1049-001, Lisbon, Portugal}

\author{Sante Carloni}
\email{sante.carloni@gmail.com}
\affiliation{Centro de Astrof\'{\i}sica e Gravita\c c\~ao - CENTRA,
Departamento de F\'{\i}sica,
Instituto Superior T\'{e}cnico - IST,
Universidade de Lisboa - UL,
Avenida Rovisco Pais 1, 1049-001, Lisbon, Portugal}

\author{Jos\'{e} P. S. Lemos}
\email{joselemos@ist.utl.pt}
\affiliation{Centro de Astrof\'{\i}sica e Gravita\c c\~ao - CENTRA,
Departamento de F\'{\i}sica,
Instituto Superior T\'{e}cnico - IST,
Universidade de Lisboa - UL,
Avenida Rovisco Pais 1, 1049-001, Lisbon, Portugal}

\begin{abstract}

Using a dynamical system approach we study the cosmological phase
space of the generalized hybrid metric-Palatini gravity theory,
characterized by the function $f\left(R,\mathcal R\right)$, where $R$
is the metric scalar curvature and $\mathcal R$ the Palatini scalar
curvature of the spacetime.  We formulate the propagation equations of
the suitable dimensionless variables that describe FLRW universes as
an autonomous system. The fixed points are obtained for four different
forms of the function $f\left(R,\mathcal R\right)$, and the behavior
of the cosmic scale factor $a(t)$ is computed. We show that due to the
structure of the system, no global attractors can be present and also
that two different classes of solutions for the scale factor $a(t)$
exist. Numerical integrations of the dynamical system equations are
performed with initial conditions consistent with the observations of
the cosmological parameters of the present state of the Universe. In
addition, using a redefinition of the dynamic variables, we are able
to compute interesting solutions for static universes.

\end{abstract}

\maketitle

%%%%%%%%%%%%%%%%%%%%%%%%%%%%%%%%%%%%%%%%%%%%%%%%%%
\section{Introduction}
%%%%%%%%%%%%%%%%%%%%%%%%%%%%%%%%%%%%%%%%%%%%%%%%%

A great number of modifications and extensions of general relativity
have been proposed to explain both inflation and the accelerated
expansion of the Universe. Among those proposals a class of theories
in which the gravitational Lagrangian contains higher order terms has
received much attention. One of the most extensively studied higher
order theories is the so-called $f\left(R\right)$-gravity theory,
where the action depends on an arbitrary function $f$ of the scalar
curvature $R$, see
e.g.~\cite{Capozziello:2007ec,Sotiriou:2008rp,DeFelice:2010aj}.

There are two approaches to obtain the field equations from the
Lagrangian of this theory. One is the metric approach, where the
metric $g_{\mu\nu}$ is considered to be the only dynamical variable in
the action. The other is the Palatini approach, where both the metric
$g_{\mu\nu}$ and the connection $\Gamma^\alpha_{\mu\nu}$ are
considered to be independent dynamical variables. Both approaches have
been used to study cosmological models which contain accelerated
expansion periods \cite{Amendola:2010bk,Carroll:2003wy,Olmo:2011uz}.

In spite of their success in reproducing the accelerated behavior of
the Universe, the metric and the Palatini approaches of
$f\left(R\right)$ present some shortcomings. In the metric approach,
the theory is not proven to be able to reproduce the observed dynamics
of objects in the solar system due to the appearance of instabilities
\cite{Dolgov:2003px,Sotiriou:2006sf}. In addition, cosmological issues
related to the background expansion
\cite{Amendola:2006kh,Amendola:2006we} and structure formation
\cite{Tsujikawa:2007tg,delaCruzDombriz:2008cp} have been pointed
out. The use of the Palatini approach can solve the problems with the
solar system dynamics, but at the same time other issues related to
physics of compact stars \cite{Reijonen:2009hi} and with the evolution
of cosmological perturbations \cite{Koivisto:2005yc,Koivisto:2006ie}
are known to appear. Since we expect any modification of general
relativity to work at multiple scales, a number of mechanisms have
been proposed to solve these problems
\cite{Koivisto:2012za,Burrage:2014uwa,Capozziello:2007eu}. See also
\cite{Nojiri:2010wj,Nojiri:2017ncd} for detailed reviews on these
matters.

A new class of modified theories of gravity that is able to solve
these difficulties has been proposed. It is called hybrid
metric-Palatini gravity.  In this class of theories a nonlinear
Palatini-like term $f\left(\mathcal R\right)$ is added to the usual
Einstein-Hilbert action to get an action with the terms
$R+f\left(\mathcal R\right)$, where $\mathcal{R}$ is the Palatini
scalar curvature defined in terms of an independent connection
\cite{Capozziello:2015lza}.  In this theory, not only cosmological
solutions consistent with both the large scale acceleration and the
solar system dynamics have been found \cite{Harko:2011nh} but also
wormhole solutions \cite{Capozziello:2012hr} and other astrophysical
and cosmological applications such as a solution for a static universe
\cite{Boehmer:2013oxa}, models for galactic rotational curves
\cite{Capozziello:2013yha}, and the virial theorem
\cite{Capozziello:2012qt}. The phase space of the hybrid
metric-Palatini gravity was studied in the Einstein frame through a
dynamical system analysis and shown to have attractors related to
exponential solutions \cite{Carloni:2015bua}. Recently, it was also
shown that the inclusion of a Higgs field in the Palatini formulation
of pure $R^2$ gravity required the use of a hybrid metric-Palatini
formalism in order to preserve Weyl invariance
\cite{PhysRevD.99.124018}.

A natural generalization of the hybrid metric-Palatini theory is to
consider that the action can be an arbitrary function of both the
scalar curvature $R$ and the Palatini curvature $\mathcal R$ as
$f\left(R,\mathcal R\right)$ \cite{Tamanini:2013ltp}. This generalized
hybrid metric-Palatini theory allows for the study of models such as
products between $R$ and $\mathcal R$ which were not covered by the
nongeneralized version of the theory.  Now, the majority of the
studies on nongeneralized and generalized hybrid metric-Palatini
theories are performed by the definition of a scalar field
representation, by which one transforms the geometrical
$f\left(R,\mathcal R\right)$ action into the action of an equivalent
scalar-tensor theory with one or two scalar fields
\cite{Tamanini:2013ltp,Rosa:2017jld} (see also
\cite{Capozziello:2015lza}). The advantage of doing so is that the
order of the field equations is reduced when the scalar fields are
introduced, thus simplifying the study of the equations of motion
\cite{Wands:1993uu}.

Whatever the representation, the cosmology of generalized hybrid
metric-Palatini theories can be efficiently analyzed using the
dynamical systems approach \cite{Wiggins,Perko}. This method consists
in the definition of
a set of specific variables by which the cosmological equations can be
converted into an autonomous system of first order differential
equations. The analysis characteristic of the phase space of this
system can then offer some semiquantitative information on the
evolution of the cosmology.  The first phase space analysis of the
scalar tensor representation of hybrid metric-Palatini theories was
performed in detail in \cite{Capozziello:2015lza}. However, both the
definition of the scalar field and the one of the dynamical system
variables, which correspond to a rearrangement of the degrees of
freedom of the theory, might hide some features of this class of
theories both at the level of the phase space and of the space of
solutions. In \cite{Carloni:2015bua}, instead, the phase space of this
theory was analyzed without introducing scalar fields.  See also
\cite{Bahamonde:2017ize} for a review of this technique in a variety
of cosmological models.

Obtaining the orbits over the entire phase-space requires in the
hybrid metric-Palatini, as in other theories, to perform a numerical
integration. To do so we need to set numerous initial
conditions. Here, these are taken from the observable cosmological
parameters, such as the scalar curvature $k$, the energy density
$\Omega$, and the derivatives of the scale factor $a$. Some of these
parameters have been measured experimentally \cite{Aghanim:2018eyx},
but others corresponding to third-order derivatives of the scale
factor, and beyond, are still poorly constrained and model dependent
\cite{Mamon:2018dxf}. In general, the orbits may approach an
attractor, but sometimes other behaviors can occur, for instance
the orbit can tend to a big rip scenario \cite{Caldwell:2003vq}.
There are other important ways to compare a
given cosmological model to observations.
For instance, supernovae Ia data have given precise
distance modulus to constrain the parameters
of the theory, for an example of how to deal with these data see
\cite{Carloni:2018ioq}.

The objective of this paper is to perform a dynamical system analysis
of the cosmology of the generalized hybrid metric-Palatini without
using the scalar-tensor representation. 
The paper is organized as follows: In Sec.~\ref{secII}, we derive
the field equations in the geometrical representation for a FLRW,
i.e., a Friedmann-Lema\^itre-Robertson-Walker,
metric, characterized by the scale factor $a(t)$, where $t$ is the
cosmological time, and define the needed variables to write the
Friedmann and the Raychaudhuri equations in a simple form. In
Sec.~\ref{secIII}, we define the dynamical variables of the system,
compute their respective dynamical equations, and show how to
obtain a given solution for $a(t)$ for a specific fixed point. In
Sec.~\ref{secIV}, we obtain the fixed points for four different
models for the function $f$. In Sec.~\ref{data} we perform
a numerical integration of one fourth-order and one second-order models
subjected to initial conditions consistent with cosmological
observations. In Sec.~\ref{Static}, we perform an analysis of the
solutions with the Hubble parameter $H$ set to zero, $H=0$, i.e.,
static universes are analyzed. In Sec.~\ref{conc}, we give the
conclusions.

%%%%%%%%%%%%%%%%%%%%%%%%%%%%%%%%%%%%%%%%%%%%%%%%%%
\section{Basic Equations}\label{secII}
%%%%%%%%%%%%%%%%%%%%%%%%%%%%%%%%%%%%%%%%%%%%%%%%%
\subsection{Fundamentals}\label{funds}

Consider the action of the generalized hybrid metric-Palatini
modified theory of gravity, given by 
\be\label{genact}
S=\frac{1}{2\kappa^2}\int_\Omega\sqrt{-g}f\left(R,\cal{R}\right)d^4x
+S_m(g_{ab}, \chi),
\ee
where $\kappa^2\equiv 8\pi G/c^4$, $G$ is the gravitational constant,
$c$ is the velocity of light, $g$ is the determinant of the metric
$g_{ab}$, $f$ is a function of $R$ and $\mathcal{R}$, and $S_m$ is the
matter action, in which matter is minimally coupled to the metric
$g_{ab}$, and $\chi$ collectively denotes the matter fields. $R$ is
the metric Ricci scalar and $\mathcal{R}\equiv g_{ab}\mathcal{R}^{ab}$
is the Palatini scalar curvature, with $\mathcal{R}^{ab}$ being
defined in terms of an independent connection $\hat\Gamma^c_{ab}$ as
\be
\mathcal{R}_{ab}=\partial_c\hat\Gamma^c_{ab}-\partial_b
\hat\Gamma^c_{ac}+\hat\Gamma^c_{cd}\hat\Gamma^d_{ab}-\hat\Gamma^c_{ad}
\hat\Gamma^d_{cb}\,.
\label{RicciPalatini}
\ee
We set $G=1/8\pi$, $c=1$, and so $\kappa^2=1$.

Varying the action in Eq.~(\ref{genact}) with respect to the metric
$g_{ab}$ and the independent connection $\hat\Gamma^c_{ab}$ yields the
following field equations:
\be\label{field1}
\frac{\partial f}{\partial R}R_{ab}+\frac{\partial f}{\partial 
\mathcal{R}}\mathcal{R}_{ab}-\frac{1}{2}g_{ab}f\left(R,\cal{R}\right)-
\left(\nabla_a\nabla_b-g_{ab}\Box\right)\frac{\partial f}{\partial R}= T_{ab},
\ee
and
\be
\hat\nabla_c\left(\sqrt{-g}\frac{\partial f}{\partial \cal{R}}g^{ab}\right)=0,
  \label{indconnnection}
\ee
respectively, where $\nabla_a$ and $\hat\nabla_a$ are the covariant derivatives of the connections $\Gamma$ and $\hat\Gamma$ respectively, $\Box$ is the d'Alembert operator, and $T_{ab}$ is the matter stress-energy tensor. The equation of motion (\ref{indconnnection}) implies that the independent connection $\hat\Gamma$ is the Levi-Civita connection of a 
new metric tensor $h_{ab}$ which is conformally related to $g_{ab}$ by 
\be\label{hgf}
h_{ab}=g_{ab} \frac{\partial f}{\partial \cal{R}}\,.
\ee
The independent connection $\hat\Gamma$ can then be written in terms of the metric $h_{ab}$ as
\be
\hat\Gamma^a_{bc}=\frac{1}{2}h^{ad}\left(\partial_b h_{dc}+
\partial_c h_{bd}-\partial_d h_{bc}\right)\,,
\ee
and the relation between $R$ and $\cal{R}$ is given by the dynamical equation
\be\label{Curvatures_Link}
\Box \left(\ln \frac{\partial f}{\partial \cal{R}}\right)+\frac{3}{2} \left(\nabla_a \ln \frac{\partial f}{\partial \cal{R}}\right)\left(\nabla^a \ln \frac{\partial f}{\partial \cal{R}}\right)+R-{\mathcal R}=0\,.
\ee
Thus, the new metric $h_{ab}$ is an auxiliary metric related to the independent connection, that was used to define the Palatini Ricci tensor, given by Eq.~(\ref{RicciPalatini}). We emphasize that matter is coupled to the physical metric $g_{ab}$, so that only the Levi-Civita connection $\Gamma(g_{ab})$  should be used in the geodesic equation applied to the metric-Palatini theory.
Note also that since the matter action $S_m$ does not depend on the connection $\hat\Gamma$, the equation of motion for this connection is independent of the stress-energy tensor, whereas the same does not happen to the equation of motion for the metric $g_{ab}$.

Using the definition of the Einstein tensor,
\be
G_{ab}= R_{ab}-\frac{1}{2}Rg_{ab},
\ee
and introducing an Einstein tensor for the Palatini field given by
\be
\mathcal{G}_{ab} =\mathcal{R}_{ab}-\frac{1}{2}\mathcal{R}g_{ab},
\ee
we can write Eq.~\eqref{field1} in a more useful way as
\be
\frac{\partial f}{\partial R}G_{ab}+\frac{\partial f}{\partial \mathcal{R}}\mathcal{G}_{ab}-\frac{1}{2}g_{ab}\left[f\left(R,\cal{R}\right)-\frac{\partial f}{\partial R}R-\frac{\partial f}{\partial \mathcal{R}}\mathcal{R}\right]-\left(\nabla_a\nabla_b-g_{ab}\Box\right)\frac{\partial f}{\partial R}= T_{ab}.
\ee
We also define the auxiliary variables $E$ and $F$ as 
\be
E\left(R,\mathcal R\right)=\frac{\partial f}{\partial R},
\ee
\be
F\left(R,\mathcal R\right)=\frac{\partial f}{\partial \mathcal{R}},
\label{f}
\ee
to obtain
\be\label{middin}
EG_{ab}+F\mathcal{G}_{ab}-\frac{1}{2}g_{ab}\left[f\left(R,\cal{R}\right)-ER-F\mathcal{R}\right]-\left(\nabla_a\nabla_b-g_{ab}\Box\right)E= T_{ab}.
\ee
We shall be working with functions $f$ that satisfy the Schwartz theorem, and therefore their crossed derivatives are equal, which is also true for the functions $E$ and $F$. This feature imposes the following constraints on the derivatives of the functions $E$ and $F$:
\be
E_\mathcal R=F_R,\ \ \ F_{\mathcal R R}=F_{R\mathcal R}=E_{\mathcal R\mathcal R},\ \ \ E_{R\mathcal R}=E_{\mathcal R R}=F_{RR},
\ee
where the subscripts $R$ and $\mathcal R$ denote derivatives with respect to $R$ and $\mathcal R$, respectively.

The set of equations derived from Eq.~\eqref{middin}  are in principle of order 4 in the metric tensor. However, there are functions $f$ for which these field equations contain only terms of order 2. This happens if the following conditions are satisfied:
\begin{align}\label{Cond2Ord}
\nonumber &F_R^2-F_\mathcal R E_R =0,\\
&F_R^2F_{\mathcal R\mathcal R}-2F_RF_\mathcal RF_{R\mathcal R}+F_\mathcal R^2F_{RR}=0,\\
\nonumber &F_\mathcal R^3E_{RR}-3F_\mathcal R^2F_RF_{RR}+3F_\mathcal RF_R^2F_{R\mathcal R}-F_R^3F_{\mathcal R\mathcal R}=0.
\end{align}
We assume these conditions hold. 
A class of functions that are solutions of the conditions~\eqref{Cond2Ord} is 
\begin{equation}
f=\alpha+\mathcal{R}\, g\left(\frac{R}{\mathcal{R}}\right)+R\,
h\left(\frac{\mathcal R}{R}\right),
\end{equation}
where $\alpha$ is a constant and
here $g$ and $h$ denote functions of their arguments.
In the following we will examine in detail a member of
this class of functions.

\subsection{The cosmological geometry}\label{flrw}

From this point on, we consider the FLRW spacetime. In
spherical comoving coordinates 
$(t,r,\theta,\phi)$
the line element can be written as
\be\label{FLRW}
ds^2=-dt^2+a^2\left(t\right)\left[\frac{dr^2}{1-kr^2}+
r^2d\theta^2+r^2\sin^2\theta d\phi^2\right],
\ee
where $a(t)$ is the scale factor, and 
$k$ spatial curvature parameter,
which can assume three values, $k=1,0,-1$, for
spherical, flat, and hyperbolic geometries, respectively.
A quantity that appears quite often is the Hubble
parameter defined by 
\be
H=\frac{\dot a}{a}\,,
\label{hubblepar}
\ee
where a dot denotes a derivative with respect to $t$.
We define an auxiliary variable ${\overline a}$ as
${\overline a}=\sqrt{F}a\left(t\right)$
and a new time variable
$\tau=\sqrt{F}t$, where $F$ is given in Eq.~(\ref{f}). 
Then a modified Hubble parameter $\mathcal H$
defined as $\mathcal{H}=\frac{\dot{\overline a}}{{\overline a}}$
is given by
\be\label{dinh}
{\mathcal H}=H+\frac{\dot F}{2F}\,.
\ee

We consider a fluid for which the  the stress-energy tensor
is of a perfect fluid,
\be\label{tab}
T^a_b=\left(-\rho,p,p,p\right)\,,
\ee
where $\rho$ is the fluid's energy density and $p$ its isotropic pressure.
We also impose an equation of state for the fluid of the form,
\be\label{eos}
p=w\rho,
\ee
where $w$ is a constant. For $w=0$ the fluid is dust, i.e., $p=0$.

With the definitions given above and the characterization of the fluid
we can write the Friedmann equation and the Raychaudhuri equation
for this system as
\be\label{dinF}
\left(\frac{\dot a}{a}\right)^2+\frac{k}{a^2}\left(1+\frac{F}{E}\right)+\frac{F}{E}\mathcal{H}^2+\frac{1}{6E}\left(f-ER-F\mathcal{R}\right)+\frac{\dot a \dot E}{aE}-\frac{\rho}{3E}=0,
\ee
\be\label{dinR}
\frac{\ddot a}{a}-\frac{F}{E}\left(\frac{k}{a^2}+\mathcal{H}^2\right)+\frac{1}{6E}\left(f-RE\right)+\frac{1}{6E}\left(\rho+3p\right)+\frac{\dot a \dot E}{2aE}+\frac{\ddot E}{2E}=0\,,
\ee
respectively. These two equations are the relevant components of
Eq.~\eqref{middin}.
Note that in defining these equations we have divided by $E$. This operation will
introduce a divergence when $E=0$. Such divergence will become relevant in the analysis.
The conservation of the stress energy tensor given by $\nabla_aT^{ab}=0$ becomes here 
\be\label{mattercons}
\dot\rho+3\frac{\dot a}{a}\left(1+w\right)\rho=0.
\ee
The equation of state Eq.~\eqref{eos}
together with the three equations
of motion Eqs.~\eqref{dinF}-\eqref{mattercons}
are the four
equations that close the system.

In cosmological models it is useful to define, besides the Hubble parameter as given in
Eq.~\eqref{hubblepar}, three other cosmological parameters, called deceleration, jerk,
and snap, as
\be\label{qdec0}
{q{}}= -\frac{\ddot{a}}{a H^2}\,,
\ee
\be\label{jerk0}
{j{}}=\frac{\dddot{a}}{a H^3}\,,
\ee
\be\label{snap0}
{s{}}=\frac{\ddddot{a}}{a H^4}\,,
\ee
respectively.

%%%%%%%%%%%%%%%%%%%%%%%%%%%%%%%%%%%%%%%%%%%%%%%%%%
\section{Dynamical System Approach}\label{secIII}
%%%%%%%%%%%%%%%%%%%%%%%%%%%%%%%%%%%%%%%%%%%%%%%%

\subsection{Equations for the dynamical system}

In dealing with dynamical systems, one must study the dimensional structure of the theory, because the number of dynamical variables and equations needed to describe the system will depend on the number of dimensional constants present in the theory. Therefore, we introduce a new nonnegative constant $R_0$ such that the quotients $R/R_0$ and $\mathcal R/R_0$ are dimensionless. In addition, we also introduce dimensionless parameters in the form of starred greek letters, such as $\alpha_*$ and so on, which will represent the product between the coupling constant of the additional invariants and a power of $R_0$. With these considerations, we can write Eq.~\eqref{genact} as
\be
S=\int_\Omega\sqrt{-g}\,f\left(\frac{R}{R_0},\frac{\mathcal{R}}{R_0},\alpha_*,...\right)d^4x+S_m(g_{\mu\nu}, \chi),
\ee
where the function $f$ retains the same properties as in the action of Eq.~\eqref{genact}. The advantage of this formalism is that instead of needing one dynamical variable for each dimensional constant, we only need a dynamical variable related to $R_0$, since the starred parameters become dimensionless.

Note that the cosmological equations, Eqs.~\eqref{dinF} and~\eqref{dinR}, depend nontrivially on time derivatives of the functions $F$ and $E$. These functions can be taken as general functions of $R$ and $\mathcal R$, so that their time derivatives can be written as functions of time derivatives of the curvature scalars, which are themselves functions of time. We therefore compute the time derivatives of $R$ and $\mathcal R$. To do so, we first define the dimensionless time variable,
\be
N=\log \left(\frac{a}{a_0}\right)\,,
\ee
where $a_0$ is some constant with dimensions of length to guarantee that the argument of the logarithm is dimensionless. We also define
\be
O'=\frac{\dot O}{H},
\ee
for any quantity $O$,
where the prime $'$ denotes a derivative with respect to $N$ and $H$ is the Hubble parameter
of Eq.~\eqref{hubblepar}.
We further redefine the 
acceleration, jerk, and snap parameters
given in Eqs.~(\ref{qdec0})-(\ref{snap0}),
as new dimensionless parameters 
${\mathcal q}, {\mathcal j},$ and ${\mathcal s}$, as
\be\label{qpar}
{\mathcal q}=\frac{H'}{H},
\ee
\be\label{jpar}
{\mathcal j}=\frac{H''}{H},
\ee
\be\label{spar}
{\mathcal s}=\frac{H'''}{H}.
\ee
Using the previous definitions, the Ricci tensor $R$ and its derivatives with respect to $t$ become
\be\label{scaric}
R=6\left[\left({\mathcal q}+2\right)H^2+\frac{k}{a^2}\right],
\ee
\be\label{dotr}
\dot R=6H\left[\left({\mathcal j}+{\mathcal q}\left({\mathcal q}+4\right)\right)H^2-\frac{2k}{a^2}\right],
\ee
\be\label{ddotr}
\ddot R=6H^2\left[\Big( {\mathcal s}+4{\mathcal j}\left(1+{\mathcal q}\right)+{\mathcal q}^2\left({\mathcal q}+8\right)\Big) H^2+2\left(2-{\mathcal q}\right)\frac{k}{a^2}\right].
\ee

Now we obtain expression for $\mathcal R$ and its derivatives.
We use again the variable ${\overline a}=\sqrt{F}a\left(t\right)$
and the time variable $\tau=\sqrt{F}t$ so that
${\overline a}^{\dagger}$ is defined as
${\overline a}^{\dagger}=\frac{\dot{\overline a}}{\sqrt{F}}$,
$\dagger$ denoting a derivative with respect to $\tau$.
Then the Palatini scalar curvature $\mathcal R$ is
$\mathcal{R}=6F\left[\frac{{\overline a}^{\dagger\dagger}}{{\overline a}}+
\left(\frac{{\overline a}^\dagger}{{\overline a}}\right)^2+\frac{k}{{\overline a}^2}\right]$
which then yields
\be\label{dinric}
\mathcal{R}=6
\left(\mathcal{\dot H}+\mathcal{H}^2+\mathcal{H}H+\frac{k}{a^2}\right).
\ee
To find expressions for the derivatives of $\mathcal{R}$ we compute the total derivative of $F$ with respect to $t$ and then use Eqs.~\eqref{dinh} and~\eqref{dinric} to solve with respect to $\mathcal {\dot R}$. We obtain
\be\label{dotmathcalr}
\mathcal{\dot R}=\frac{1}{F_\mathcal{R}}\left[\left(\mathcal H-H\right)2F-F_R \dot R\right],
\ee
\beq
&&\mathcal{\ddot R}=-\frac{1}{F_\mathcal R^2}\left(F_{\mathcal R R}\dot R+F_{\mathcal R \mathcal R}\mathcal{\dot R}\right)\left[\left(\mathcal H-H\right)2F-F_R\dot R\right]+\frac{2F}{F_R}\left(\frac{\mathcal R}{6}-\mathcal H^2-\mathcal H H-\frac{k}{a^2}-{\mathcal q}H^2\right)+\nonumber \\
&&+\frac{1}{F_\mathcal R}\left[2\left(\mathcal H-H\right)\left(F_R \dot R + F_\mathcal{R} \mathcal{\dot R}\right)-F_R\ddot R-\dot R\left(F_{RR}\dot R+F_{R\mathcal R}\mathcal{\dot R}\right)\right]\,,
\eeq
where $\dot R$, $\ddot R$, and $\mathcal{\dot R}$
have already been computed in Eqs.~\eqref{dotr},~\eqref{ddotr}, and~\eqref{dotmathcalr}, respectively.
These results completely determine the forms of
the first and second time derivatives of $F$ and $E$.

Let us now define a set of dynamical dimensionless variables as
\be\label{dynvar}
K=\frac{k}{a^2H^2},\ \ \ X=\frac{\mathcal H}{H},\ \ \ Y=\frac{R}{6H^2}, \ \ \ Z=\frac{\mathcal R}{6H^2}, \ \ \ Q={\mathcal q}, \ \ \ J={\mathcal j}, \ \ \ S={\mathcal s},\ \ \ \Omega=\frac{\rho}{3H^2E},\ \ \ A=\frac{R_0}{6H^2}.
\ee
For consistency of notation, let us also redefine ${\mathcal s}$ as
\be\label{saS}
S={\mathcal s}\,.
\ee
Note that $S$ is not a dynamic variable. This
is because since the theory is fourth-order,
$S$ can be obtained from the field equations,
and we do not need to write a dynamical equation for it.
The Jacobian $J$ of the definition of variables~\eqref{dynvar}
can be written in the form,
\be\label{jacobian}
J=\frac{1}{108a^2H^9E},
\ee
which means that it has a different form for each choice of the function $f$. In order to guarantee that the variables in Eq.~\eqref{dynvar} cover the entire phase space of the cosmological equations, i.e., they constitute a global set of coordinates for it, $J$ must always be regular, i.e.,
finite and nonzero, $J\neq0,\infty$. When the Jacobian is not regular the definition in Eq.~\eqref{dynvar} is not invertible, and therefore there can be features of the field equations which are not preserved in the phase space of Eq.~\eqref{dynvar} and features of the  phase space of Eq.~\eqref{dynvar} which are spurious, including fixed points. From Eq.~\eqref{jacobian} it is evident that a regular Jacobian corresponds to $E\neq0,\infty$.
The case $J=0$, $E=\infty$, corresponds to a true singularity in Eqs.~(\ref{dinF})-(\ref{dinR}) as well as
in Eq.~\eqref{middin}.  The case $J=\infty$, $E=0$, instead, corresponds to a singularity for Eqs.~(\ref{dinF})-(\ref{dinR}) but not for Eq.~\eqref{middin}. This implies that the solutions of Eq.~\eqref{middin} associated to $E=0$ will not be represented in the phase space.  
In the following the fixed points for which $J=0,\infty$  will not be included in our analysis unless they are attractors in the phase space. The only exception to this choice will be the fixed points representing static universe solutions which we will consider later. We will see that these points have $E=0$, but it is easy to prove via Eq.~\eqref{middin} that they  represent true solutions for the field equations.

We also define a set of auxiliary dimensionless functions as
\beq\label{GenFunct}
&&\textbf{A}=\frac{F}{E},\ \ \ \textbf{B}=\frac{f}{6EH^2},\ \ \ \textbf{C}=\frac{F_R}{F_\mathcal R},\ \ \ \textbf{D}=\frac{F}{3H^2F_\mathcal R},\ \ \ \textbf{E}=\frac{3H^2F_{R\mathcal R}}{F_\mathcal R},\nonumber \\
&&\textbf{F}=\frac{3H^2F_{RR}}{F_\mathcal R},\ \ \ \textbf{G}=\frac{3H^2F_{\mathcal R\mathcal R}}{F_\mathcal R},\ \ \ \textbf{H}=\frac{3H^2E_{RR}}{F_\mathcal R},\ \ \ \textbf{I}=\frac{E_R}{F_\mathcal R}.
\eeq
These definitions allow us to rewrite the cosmological equations, Eqs.~\eqref{dinF} and~\eqref{dinR}, as
\be\label{const1}
1-Y+\textbf{B}+K+\textbf{A}\left[K+X^2+2\textbf{C}\left(X-1\right)-Z\right]+\frac{2\textbf{A}}{\textbf{D}}\left(\textbf{I}-\textbf{C}^2\right)\left[J+Q\left(Q+4\right)-2K\right]-\Omega=0\,,
\ee
\beq
&&1+Q-Y+\textbf{B}+\frac{1+3w}{2}\Omega+\textbf{A}\left\{-\left(K+X^2\right)+\textbf{C}\left[Z-\left(X^2+1\right)-K-Q\right]+\right.\nonumber\\
&&\left.+\frac{\textbf{I}-\textbf{C}^2}{\textbf{D}}\left[J\left(5+4Q\right)+Q\left(4+9Q+Q^2\right)+2K\left(1-Q\right)+S\right]+\right.\nonumber\\
&&\left.+2\left(X-1\right)\left[\left(\textbf{E}\textbf{D}-\textbf{G}\textbf{C}\textbf{D}+\textbf{C}\right)\left(X-1\right)+2\left(J+Q\left(Q+4\right)-2K\right)\left(\textbf{F}+\textbf{G}\textbf{C}^2-2\textbf{E}\textbf{C}\right)\right]+\right.\nonumber\\
&&\left.+\frac{2}{\textbf{D}}\left[J+Q\left(Q+4\right)-2K\right]^2\left(\textbf{H}-3\textbf{C}\textbf{F}+3\textbf{C}^2\textbf{E}-\textbf{C}^3\textbf{G}\right)\right\}=0,\label{bigequation}
\eeq
respectively, and also to rewrite the definitions of $R$ and $\mathcal R$ given by Eqs.~\eqref{scaric} and~\eqref{dinric} as
\be\label{defric}
Y=K+Q+2,
\ee
\be\label{constrZ}
Z=\frac{\dot{\mathcal H}}{H^2}+X\left(X+1\right)+K,
\ee
respectively. The derivatives with respect to the dimensionless time variable $N$ of these variables become
\beq\label{SysGen}
K'&=&-2K\left(Q+1\right),\nonumber \\
X'&=&Z-X\left(X+1+Q\right)-K, \nonumber \\
Y'&=&J+Q\left(Q+4\right)-2K-2YQ, \nonumber \\
Z'&=&\textbf{D}\left(X-1\right)+\textbf{C}\left[2K-J-Q\left(Q+4\right)\right]-2ZQ,\label{system1} \\
Q'&=&J-Q^2, \nonumber \\
J'&=&S-QJ,\nonumber \\
\Omega'&=&-\Omega\left\{3\left(1+3w\right)+2Q+2\textbf{A}\left[\textbf{C}\left(X-1\right)+\frac{\textbf{I}-\textbf{C}^2}{\textbf{D}}\left(J+Q\left(Q+4\right)-2K\right)\right]\right\},\nonumber \\
A'&=&-2AQ,\nonumber
\eeq
where $S=\frac{H'''}{H}$, see Eqs.~(\ref{spar}) and (\ref{saS}),
can be written as
\beq\label{sdevel}
S&=&\frac{\textbf{D}}{\textbf{I}-\textbf{C}^2}\left\{-\frac{1}{\textbf{A}}\left(1+Q-Y+\textbf{B}+\frac{1+3w}{2}\Omega\right)+\left(K+X^2\right)-\textbf{C}\left[Z-\left(X^2+1\right)-K-Q\right]\right.-\nonumber \\
&&\left.-2\left(X-1\right)\left[\left(\textbf{E}\textbf{D}-\textbf{G}\textbf{C}\textbf{D}+\textbf{C}\right)\left(X-1\right)+2\left(J+Q\left(Q+4\right)-2K\right)\left(\textbf{F}+\textbf{G}\textbf{C}^2-2\textbf{E}\textbf{C}\right)\right]-\right. \\
&&\left.-\frac{2}{\textbf{D}}\left[J+Q\left(Q+4\right)-2K\right]^2\left(\textbf{H}-3\textbf{C}\textbf{F}+3\textbf{C}^2\textbf{E}-\textbf{C}^3\textbf{G}\right)\right\}-J\left(5+4Q\right)-Q\left(4+9Q+Q^2\right)-2K\left(1-Q\right).\nonumber
\eeq

Now, Eqs.~\eqref{const1} and ~\eqref{defric} allow us to eliminate two variables from the system. For simplicity, we chose to eliminate $Q$ and $J$, leaving a simplified system of the form,
\beq\label{simpsys}
K'&=&2K\left(K-Y+1\right),\nonumber \\
X'&=&Z-X\left(X+Y-1\right)+K\left(X-1\right),\nonumber \\
Y'&=&2Y\left(2+K-Y\right)+\frac{\textbf{D}}{2\textbf{A}\left(\textbf{C}^2-\textbf{I}\right)}\left\{1+\textbf{B}+K-Y+\textbf{A}\left[K+2\textbf{C}\left(X-1\right)+X^2-Z\right]-\Omega\right\}, \\
Z'&=&\frac{1}{2\textbf{A}\left(\textbf{C}^2-\textbf{I}\right)}\left\{4\textbf{A}\textbf{C}^2\left(2+K-Y\right)-\right.\nonumber \\
&&\left.-2\textbf{A}\textbf{I}\left[\textbf{D}\left(X-1\right)+2Z\left(2+K-Y\right)\right]-\textbf{C}\textbf{D}\left[1+\textbf{B}+K-Y-\Omega+\textbf{A}\left(K+X^2-Z\right)\right]\right\},\nonumber\\
\Omega'&=&-\Omega\left[-2+3w-\textbf{B}-3\left(K-Y\right)-\textbf{A}\left(K+X^2-Z\right)+\Omega\right]\nonumber \\
A'&=&2A\left(2+K-Y\right).\nonumber
\eeq
In the above system we have implemented the constraints given in
Eqs.~\eqref{const1} and~\eqref{defric} to keep the equation to a
manageable size. The implementation of the constraints introduces some
nontrivial structural changes in the system, like the cancellations of
the divergences.

In the following we will use the above Eqs.~\eqref{simpsys} to
explore the phase spaces of models with a given form of the function
$f(R,\mathcal R)$. We will apply the general method presented above to
a number of different functions $f$. Some of the models are chosen for
their simplicity and the analogy with some interesting $f(R)$ gravity
theories. These are
$R^n\mathcal R^m$,
 $\alpha R^n+\beta\mathcal R^m$,
$\exp\left(\frac{R}{\mathcal R}\right)$.
Others, as $R \exp\left(\frac{R}{\mathcal R}\right)$, are chosen for
the special form assumed by their field equations and the connection
with the work done in \cite{Rosa:2017jld}.

\subsection{Solution associated to a fixed point}

Before we delve into specific examples of application of the above
formalism, we can give the general solution associated to a given
fixed point. Such a solution can be found by computing the value of $S$
in Eq.~(\ref{sdevel}) using the values of the dynamic variables and
functions at that given fixed point.  At the fixed point $S$ is a
constant.  Then, $S=\frac{H'''}{H}$, defined in Eqs.~(\ref{spar}) and
(\ref{saS}), becomes a differential equation for the scale factor
$a(t)$. The equation has two possible forms one for $S=0$, the other
for $S\neq 0$.

\begin{figure}[h]
\centering 
\includegraphics[scale=0.7]{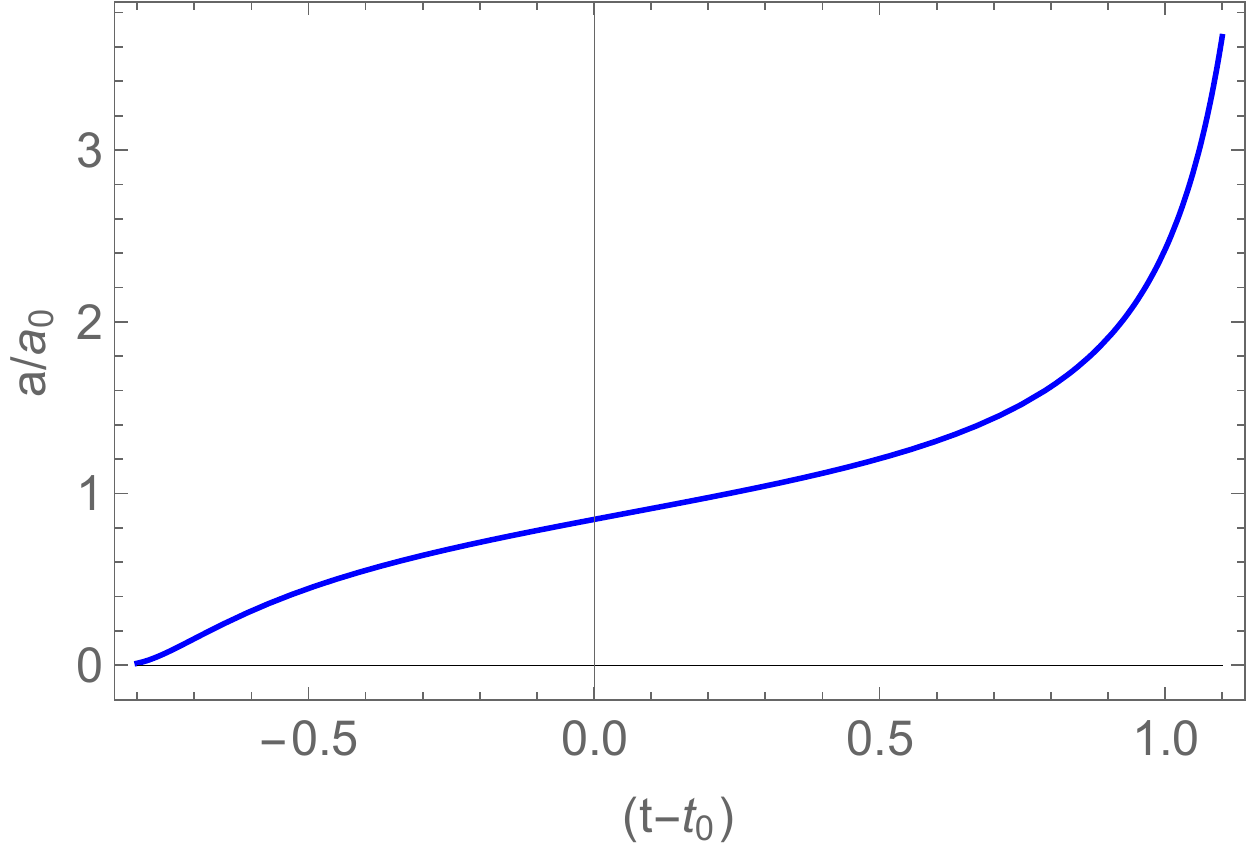}
\caption{Scale factor $a\left(t\right)$
for $S= 0$, see
Eq.~\eqref{numsca}  or Eq.~\eqref{anasca}, 
for $4H_2H_0-H_1^2>0$
with $H_0=H_1=H_2=1$, $a_0=1$, $t_0$=0.}
\label{scalefactor2}
%\vskip 0.1cm
\end{figure}

For $S=0$ the equation is
\be\label{numsca}
\frac{\dot a}{a}=H_0+H_1\ln \left(\frac{a}{a_0}\right)+
H_2\left(\ln \left(\frac{a}{a_0}\right)\right)^2,
\ \ \ \ \
\hskip 6.05cm S=0\,.\
\ee
Equation~\eqref{numsca} for $S=0$ can be solved analytically and the result is
\be\label{anasca}
a\left(t\right)=a_0\exp\left(-\frac{H_1}{2H_2}\right)
\exp\left\{\frac{\sqrt{4H_0H_2-H_1^2}}
{2H_2}\tan\left[\frac{\sqrt{4H_0H_2-
H_1^2}}{2}\left(t-t_0\right)\right]\right\},
\ \ \ \ \
\hskip 0.40cm S=0\,,
\ee
where $t_0$ is a constant representing some free initial time.  This
solution for $S=0$ will have three different behaviors depending on
the values of the integration constants $H_i$.  For the case $S=0$
with $4H_2H_0-H_1^2>0$ the scale factor $a(t)$ behavior is displayed
in Fig.~\ref{scalefactor2}, where it is seen that a finite time
singularity will appear, see also Eq.~\eqref{numsca} or
Eq.~\eqref{anasca}.  For the case $S=0$ and $4H_2H_0-H_1^2=0$, the
solution for the scale factor $a(t)$ will be a constant, and it is not
physically relevant, see Eq.~\eqref{anasca}.  For the case $S=0$ with
$4H_2H_0-H_1^2<0$ the scale factor $a(t)$ behavior is such that the
solution does not have a starting point in time, i.e., it does not
present a big bang, it is asymptotically static to the past with a
finite scale factor, and it is asymptotically zero to the future, see
Eq.~\eqref{numsca} or Eq.~\eqref{anasca}. This solution does not have
any physical relevance given the observational fact that the Universe
is undergoing a period of accelerated expansion.  The presence of
attractors with this character therefore might be a sign of a
potential instability of the model for a certain set of initial
conditions.

\newpage

\begin{figure}[h]
\centering 
\includegraphics[scale=0.7]{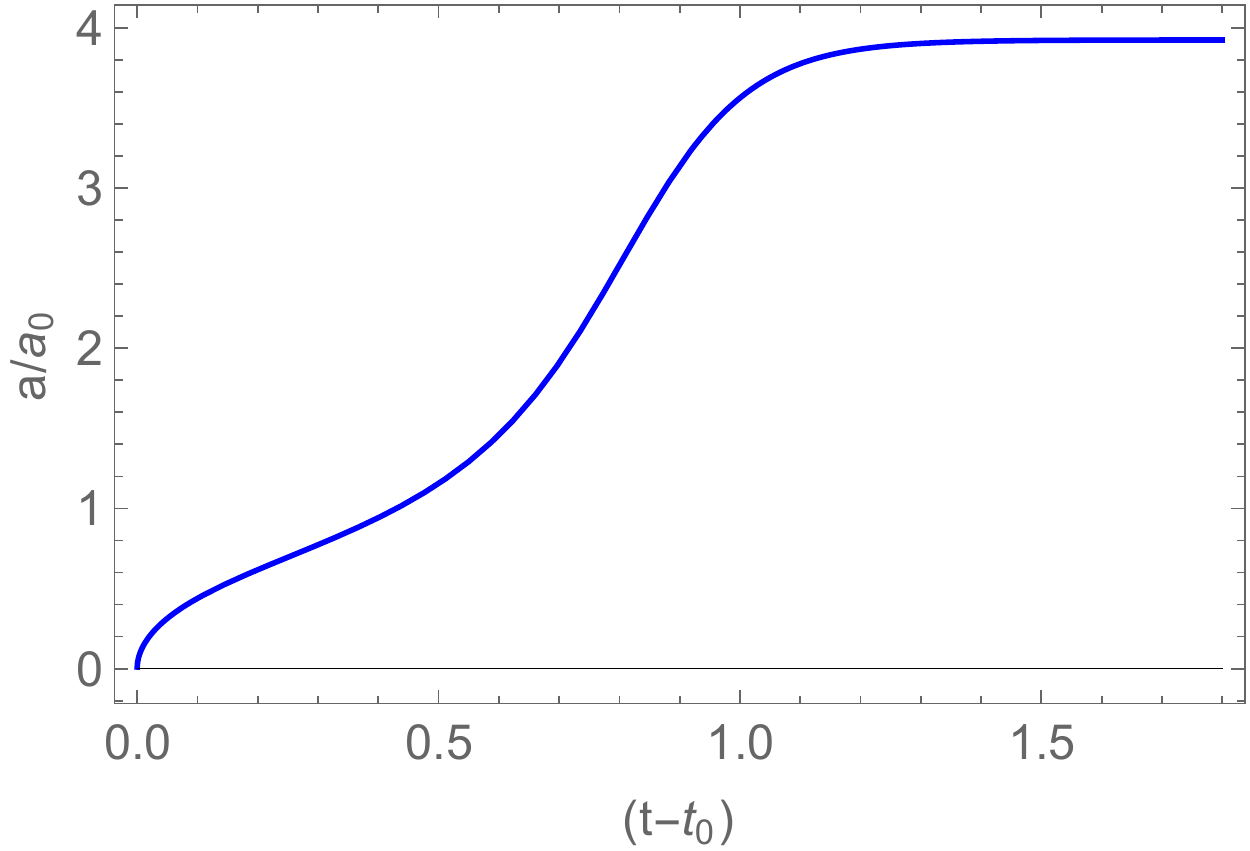}
\caption{Scale factor $a\left(t\right)$
for $S\neq 0$, see
Eq.~\eqref{numsca1} with $H_0=H_1=H_2=1$, $S=-8$ so $p=2$, $a_0=0.01$, $t_0$=0.}
\label{scalefactor1}
%\vskip 0.1cm
\end{figure}

For $S\neq0$ the equation is
\be
\frac{\dot a}{a}=H_0\left(\frac{a}{a_0}\right)^{-p}
+\left(\frac{a}{a_0}\right)^{\frac{p}{2}}\left[
H_1\sin\left(\frac{p\sqrt{3}}{2}\ln \left(\frac{a}{a_0}\right)\right)+
H_2\cos\left(\frac{p\sqrt{3}}{2}\ln \left(\frac{a}{a_0}\right)\right)
\right],\ \ \ \ \ S\neq 0\label{numsca1},
\ee
where $p=-\sqrt[3]{S}$, $H_0$, $H_1$, and $H_2$ are constants of integration, and 
$a_0$ is some constant with dimensions of length.
Note that Eq.~\eqref{numsca} for $S=0$ can be obtained
from the limit $S\to0$ of Eq.~\eqref{numsca1}
with some reworking of the constants $H_0$, $H_1$, and $H_2$.
Equation~\eqref{numsca} for $S\neq0$ can be solved numerically only.  We
plot in Fig.~\ref{scalefactor1} the behavior of the scale factor
$a(t)$ for the case $S\neq0$, where it is seen that the time evolution
of the solution will approach a constant value of the scale factor,
see also Eq.~\eqref{numsca1}. 

The two possible physical relevant
solutions, namely $S=0$ with $4H_2H_0-H_1^2>0$ and $S\neq0$, have
clearly crucial differences, the former develops a finite time
singularity, i.e.,  a big rip, 
whereas the latter is asymptotically constant.

%\newpage
%\centerline{}
\newpage
%%%%%%%%%%%%%%%%%%%%%%%%%%%%%%%%%%%%%%%%%%%%%%%%%%
\subsection{An $S=0$ analytical solution consistent
with measured cosmological parameters}
%%%%%%%%%%%%%%%%%%%%%%%%%%%%%%%%%%%%%%%%%%%%%%%%%

It is interesting to find an analytical solution
with zero snap parameter, i.e., ${s{}}=0$ or $S=0$,
consistent with measured cosmological parameters. 
According to the standard model of cosmology together with the
inflationary model, our Universe started from a big bang, suffered an
initial period of accelerated expansion called inflation, then
decelerated during the radiation and matter domination eras, and
afterwards resumed an accelerated expansion period when dark energy
became dominant over the remaining contributions to the density
parameter.  An analysis of the $S=0$ solution given in
Eq.~\eqref{anasca} and plotted in Fig.~\ref{scalefactor2} shows that
this solution qualitatively presents all these behaviors, with the
exception that the standard model of cosmology does not predict a
finite-time singularity. This is an indication that the solution in
Eq.~\eqref{anasca} might be of cosmological interest. The constants of
integration $H_i$ can be tuned in such a way that the solution
reproduces the observed values for the Hubble parameter $H$ and the
deceleration parameter ${q{}}$, and also provides a prediction for
the jerk parameter ${j{}}\left(t\right)$ given by
Eqs.~\eqref{qdec0}-\eqref{jerk0}.

We use
the solution for $a(t)$ given in 
Eq.~\eqref{anasca} which is analytic and valid for $s=0$.
The moment of the big bang corresponds to the time for which the scale
factor vanishes. For simplicity, we set
$t_0=\frac{\pi}{\sqrt{4H_0H_2-H_1^2}}$, in such
a way that the big bang occurs at the instant $t=0$. This corresponds
simply to a translation in time of the solution. The scale factor
$a\left(t\right)$ found in Eq.~\eqref{anasca},
can then be used to find
the Hubble parameter given in 
Eq.~\eqref{hubblepar},
the acceleration parameter 
given in 
Eq.~\eqref{qdec0}, and the jerk parameter
given in 
Eq.~\eqref{jerk0}. The scale factor and the
cosmological parameters then become 
\be
a\left(t\right)=a_0\exp\left(-\frac{H_1}{2H_2}\right)
\exp\left[\frac{\sqrt{4H_0H_2-H_1^2}}
{2H_2}\tan\left(\frac{\sqrt{4H_0H_2-
H_1^2}}{2}t-\frac{\pi}{2}\right)\right],
\ee
\be
H\left(t\right)=\frac{4H_0H_2-
H_1^2}{4H_2}
\frac{1}{\cos^2\left(\frac{\sqrt{4H_0H_2-H_1^2}}{2}
t-\frac{\pi}{2}\right)},
\label{hmodel}
\ee
\be
{q{}}\left(t\right)=\frac{2H_2}{\sqrt{4H_0H_2-H_1^2}}
\cos\left(
\frac{\sqrt{4H_0H_2-H_1^2}}{2}t-\frac{\pi}{2}\right)
-1,
\label{qmodel}
\ee
\beq
{j{}} \left(t\right) =&&  1+ \frac{H_2}{4H_0H_2-H_1^2}
\left[ 6H_2+
4H_2 \sin\left(\frac{\sqrt{4H_0H_2-H_1^2}}{2}t-\frac{\pi}{2}\right)+\right.
 \nonumber\\
&&
\left. 2H_2\sin\left(\sqrt{4H_0H_2-H_1^2}t-\frac{\pi}{2}\right)-
6\sqrt{4H_0H_2-H_1^2}
\cos\left(\frac{\sqrt{4H_0H_2-H_1^2}}{2}t-\frac{\pi}{2}\right)\right].
\label{jmodel}
\eeq

\begin{figure}[h]
\includegraphics[scale=0.65]{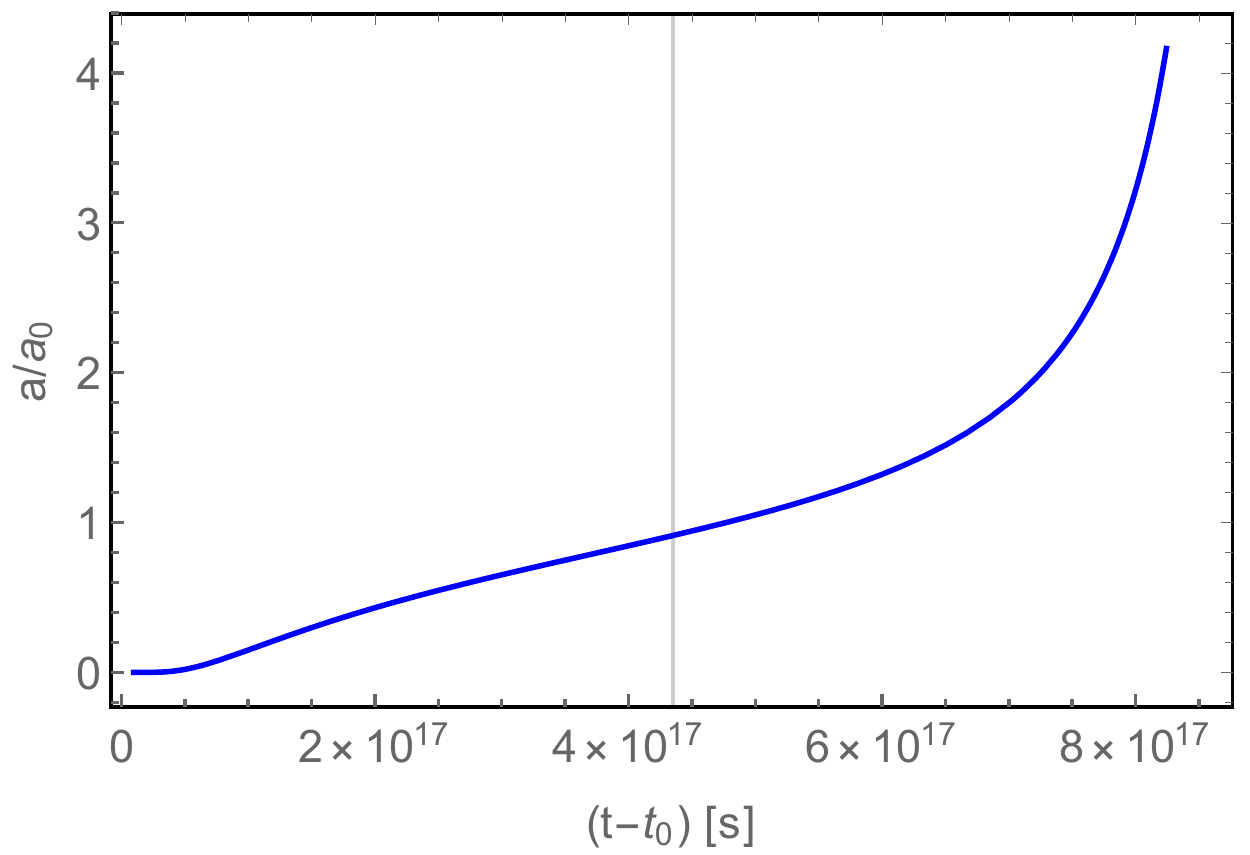}
\includegraphics[scale=0.7]{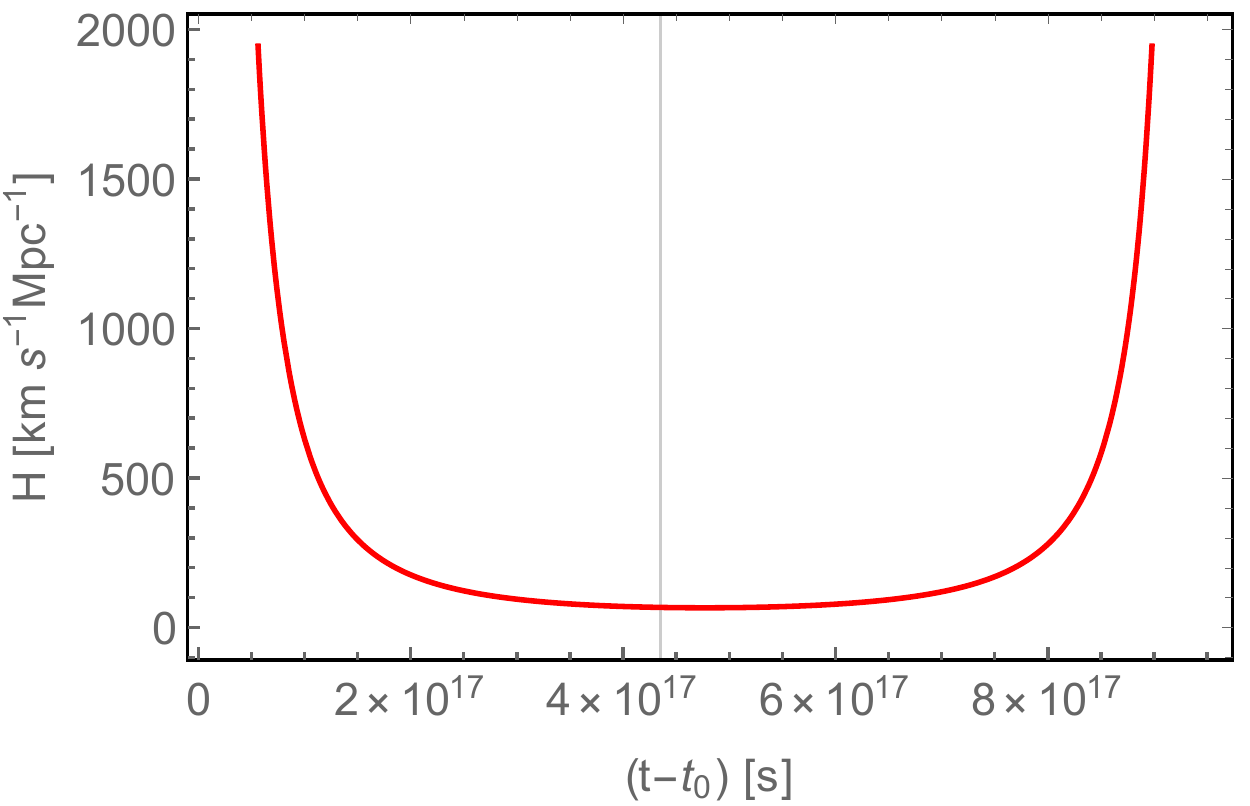}
\includegraphics[scale=0.7]{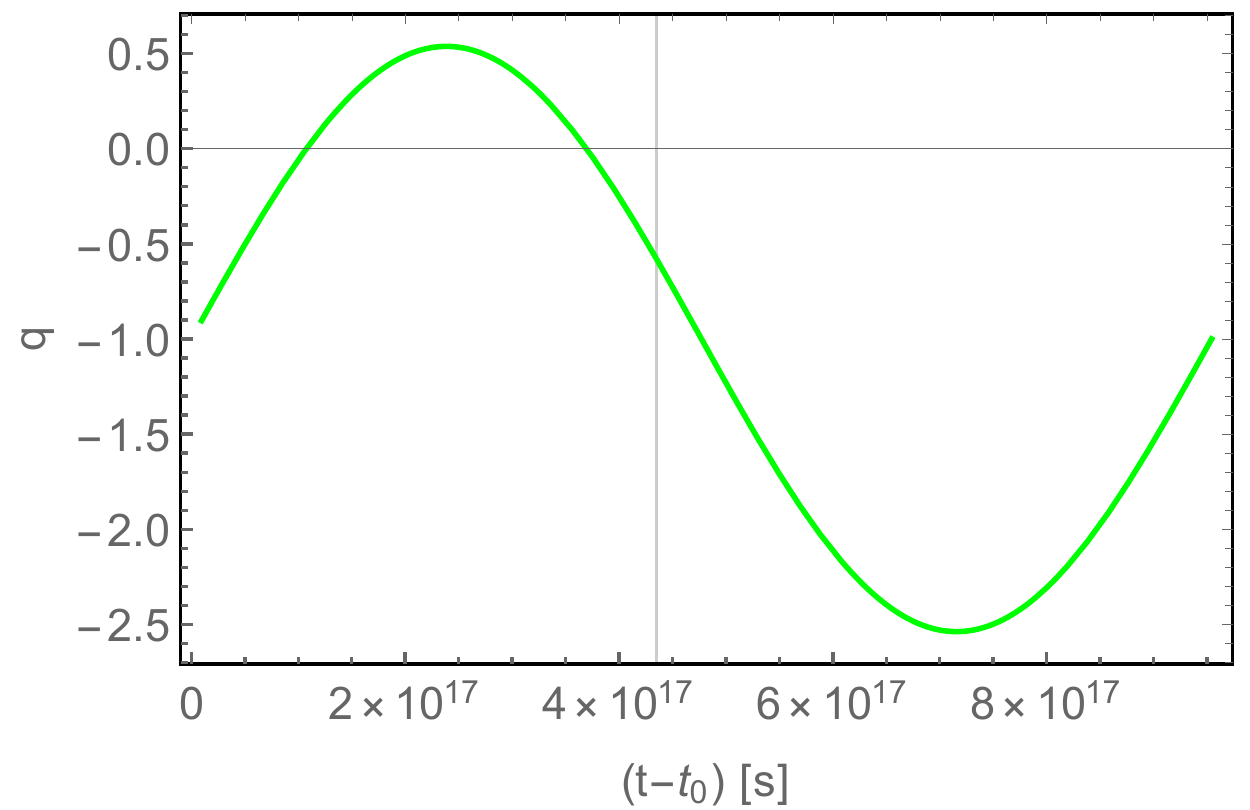}
\includegraphics[scale=0.65]{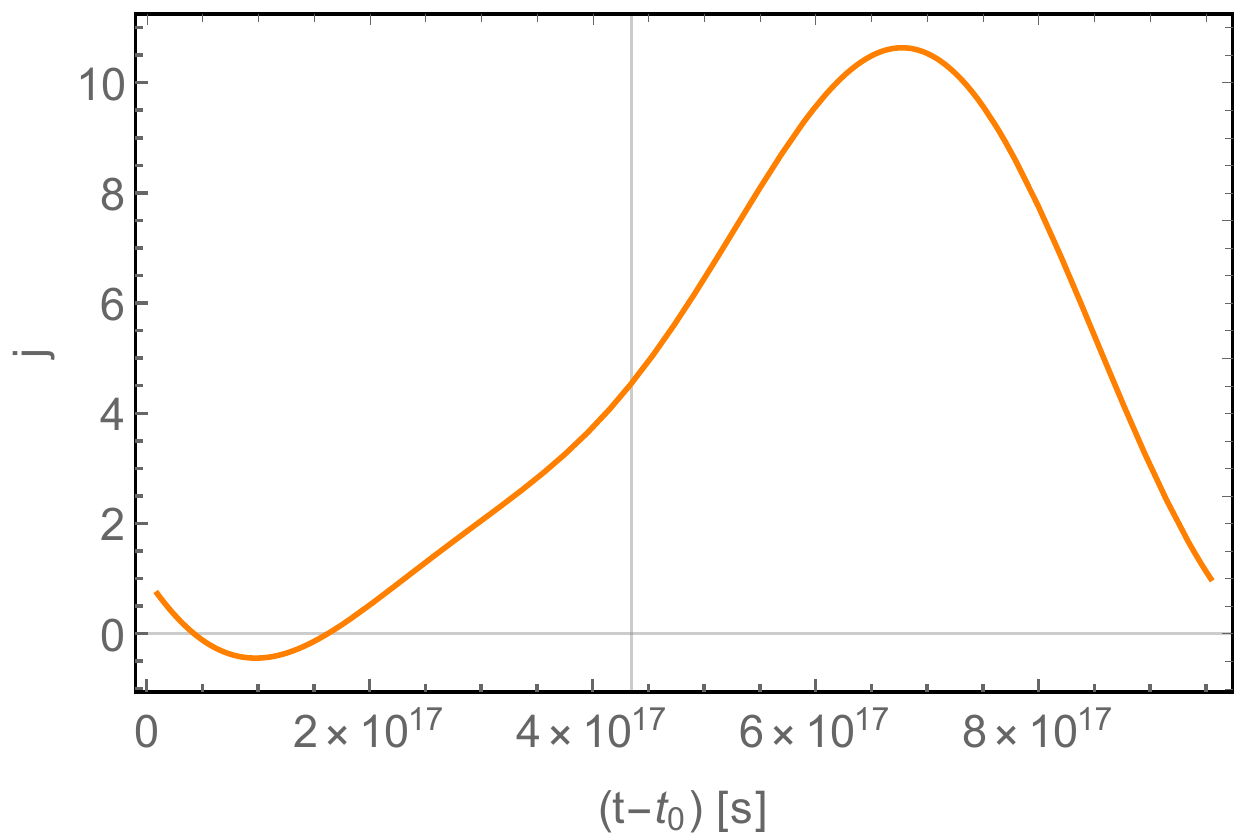}
\caption{Scale factor $a\left(t\right)$ (blue, upper left), Hubble
parameter $H\left(t\right)$ (red, upper right), deceleration parameter
${q{}}\left(t\right)$ (green, lower left), and jerk parameter $\bar
j\left(t\right)$ (orange, lower right), for the solution satisfying
the observational constraints on $H_p$ and ${q{}}_p$. The vertical
lines represent the present time.}  \label{cosmopar}
%\vskip 1cm
\end{figure}

The present values for
$H$ and ${q{}}$, which
we denote by $H_p$ and ${q{}}_p$ respectively (and not by
$H_0$ and $q_0$ as it is usual in the literature because
we have defined other parameters with such labels), have been measured
experimentally and are $H_p\sim 67.4\
\text{km}\text{s}^{-1}\text{Mpc}^{-1}\sim 2.19\times 10^{-18}\
\text{s}^{-1}$, with a relative uncertainty of less than $1\%$, and
${q{}}_p\sim -0.55$. Also, the age of the Universe, or the time
passed since the big bang, has also been measured and has a value of
$t_p\sim 13.787 \times 10^9 \text{y}\sim 4.35 \times 10^{17}\text{s}$.
Inserting $t_p$
in Eqs.~\eqref{hmodel}
and~\eqref{qmodel}, putting 
$H\left(t_p\right)=H_p$ and $\bar
q\left(t_p\right)={q{}}_p$, and using the
known values of $H_p$ and ${q{}}_p$
yields a system of two equations for
the two unknowns, namely, 
 $\sqrt{4H_2H_0-H_1^2}$
and $2H_2$, which can be solved to give
$\sqrt{4H_2H_0-H_1^2}\sim 6.584\times 10^{-18}\text{s}^{-1}$
and
$2H_2\sim
1.012\times 10^{-17}\ \text{s}^{-1}$.
These values can then be inserted in the jerk parameter
${j{}}$ of 
Eq.~\eqref{jmodel}
 to
predict the present value of the cosmological jerk parameter,
giving ${j{}}(t_p)={j{}}_p\sim 4.47$, i.e., a value
not much greater than one.
One can also compute the expected time for
the finite
time singularity to occur, which is given by
$t_s=\frac{2\pi}{\sqrt{4H_0H_2-H_1^2}}\sim 9.54\times 10^{17}\
\text{s}\sim 2.21\ t_p$. In Fig.~\ref{cosmopar} we plot
$a\left(t\right)$, $H\left(t\right)$, ${q{}}\left(t\right)$ and
${j{}}\left(t\right)$ for this model. The model does not provide a
prediction for the snap parameter ${s{}}\left(t\right)$ as we are
considering a solution with $S=0$ at a given fixed point. This
corresponds to an approximation to the real solution.

%%%%%%%%%%%%%%%%%%%%%%%%%%%%%%%%%%%%%%%%%%%%%%%%%%
\section{Examples}\label{secIV}
%%%%%%%%%%%%%%%%%%%%%%%%%%%%%%%%%%%%%%%%%%%%%%%%%
\subsection{The case of $R^n\mathcal R^m$ gravity}

In this section we consider that the
function $f$ has the form
$f=\alpha{R^n}{\mathcal R^m}$, for some constant $\alpha$
and free exponents $n$ and $m$
which can be put in the form
$f=\alpha_* \frac{R^n}{R_0^{n}}
\frac{\mathcal R^m}{R_0^{m}}$,
with $\alpha_*$ and $R_0$ constants,
and so the action $S^*$, say, is
$S^*=\int\sqrt{-g}\, \alpha_*  \frac{R^n}{R_0^{n}}
\frac{\mathcal R^m}{R_0^{m}} d^4x + {S}_m^*$, with
${S}_m^*$ the matter action.
Note that in this case $\alpha_*$ can be factored out of the action
without loss of generality by defining ${S}_m=\alpha_*^{-1}{S}_m^*$, so that
\be\label{act1}
S=\int\sqrt{-g}\,  \frac{R^n}{R_0^{n}}
\frac{\mathcal R^m}{R_0^{m}} d^4x + S_m\,.
\ee
As a consequence, there will be no need for the variable $A$ associated to the constant $R_0$,
which means that this is a degenerate case, much in the same way of the case $f(R)=R^n$
studied in \cite{Carloni:2015bua}.
The Jacobian given in Eq.~\eqref{jacobian} for this case can be written in terms of the dynamic variables and parameters as
\be
J=\frac{Y^{1-n}Z^{-m}}{n 2^{1+n+m}3^{2+n+m}H^{7+2\left(n+m\right)}a^2}.
\ee
For this Jacobian to be finite, we must exclude the value $n=0$ from the analysis and also constrain our results for the fixed points to have values for the variables $Y$ and $Z$ different from zero.

The dynamical functions in Eq.~\eqref{GenFunct} in this case are
\beq\label{GenFuncta}
&&\textbf{A}=\frac{m Y}{n Z},\ \ \ \textbf{B}=\frac{Y}{n},\ \ \ \textbf{C}=\frac{n Z}{\left(m-1\right)Y},\ \ \ \textbf{D}=\frac{2Z}{m-1},\ \ \ \textbf{E}=\frac{n}{2Y},\nonumber \\
&&\textbf{F}=\frac{n\left(n-1\right)Z}{2\left(m-1\right)Y^2},\ \ \ \textbf{G}=\frac{m-2}{2Z},\ \ \ \textbf{H}=\frac{n\left(n-1\right)\left(n-2\right)Z^2}{2m\left(m-1\right)Y^3},\ \ \ \textbf{I}=\frac{n\left(n-1\right)Z^2}{m\left(m-1\right)Y^2},
\eeq
and, once the constraints~\eqref{defric} and~\eqref{constrZ} are implemented, the dynamical system from Eq.~\eqref{simpsys} becomes 
\beq
K'&=&2K\left(K-Y+1\right),\nonumber \\
X'&=&Z-X\left(X+Y-1\right)+K\left(X-1\right),\nonumber \\
Y'&=&Y\left\{2\left(2+K-Y\right)+\frac{\left(m-1\right)}{n+m-1}\left\{1+Y\left(\frac{1}{n}-1\right)+K+\frac{m Y}{n Z}\left[K+2\frac{n Z}{\left(m-1\right)Y}\left(X-1\right)+X^2-Z\right]-\Omega\right\}\right\}, \nonumber\\
Z'&=&\frac{\left(m-1\right)}{2\left(n+m-1\right)}\left\{4nm \left(2+K-Y\right)-\right.\label{system1a} \\
&&\left.-2\left(n-1\right)\left[\frac{2Z}{m-1}\left(X-1\right)+2Z\left(2+K-Y\right)\right]-\frac{2n Z}{\left(m-1\right)}\left[1+Y\left(\frac{1}{n}-1\right)+K-\Omega+\frac{m Y}{n Z}\left(K+X^2-Z\right)\right]\right\},\nonumber\\
\Omega'&=&-\Omega\left[-2+3w-\frac{Y}{n}-3\left(K-Y\right)-\frac{m Y}{n Z}\left(K+X^2-Z\right)+\Omega\right]\,.\nonumber 
%\\ A'&=&2A\left(2+K-Y\right).\nonumber
\eeq
The set of equations given in~\eqref{system1a}  presents divergences for specific values of the parameters  for $n=0$ or $m=1$ and for any $n+m=1$, which implies that our formulation is not valid for these cases. Indeed, when this is the case the functions in~\eqref{GenFuncta} are divergent, and the analysis should be performed starting again from the cosmological
equations given by Eqs.~\eqref{dinF} and~\eqref{dinR}. The dynamical system also presents some divergences for  $Y=0$ and $Z=0$, which are due to the  very structure of the  gravitation field equations for this choice of the action. Because of these singularities the dynamical system is not $C(1)$ in the entire phase space, and one can use the standard analysis tool of the phase space only when $Y, Z\neq0$. We will pursue this kind of analysis here.

The system also presents the $K=0$ and $\Omega=0$ invariant submanifolds together with the invariant submanifold $Z=0$. The presence of the latter submanifolds allows us to solve partially the problem about the singularities in the phase space. Indeed  the presence of the $Z=0$  submanifold implies that no orbit will cross this surface. However the issue remain for the $Y=0$ hypersurface. The presence of this submanifold also prevents the presence of a global attractor  for this case. Such an attractor should have  $Z=0$, $K=0$ and $\Omega=0$ and therefore would correspond to a singular state for the theory. This feature also allows us to discriminate sets of initial conditions and of parameters values which will lead to a given time-asymptotic state for the system.

\begin{figure}[!t]
\centering
\includegraphics[scale=0.5]{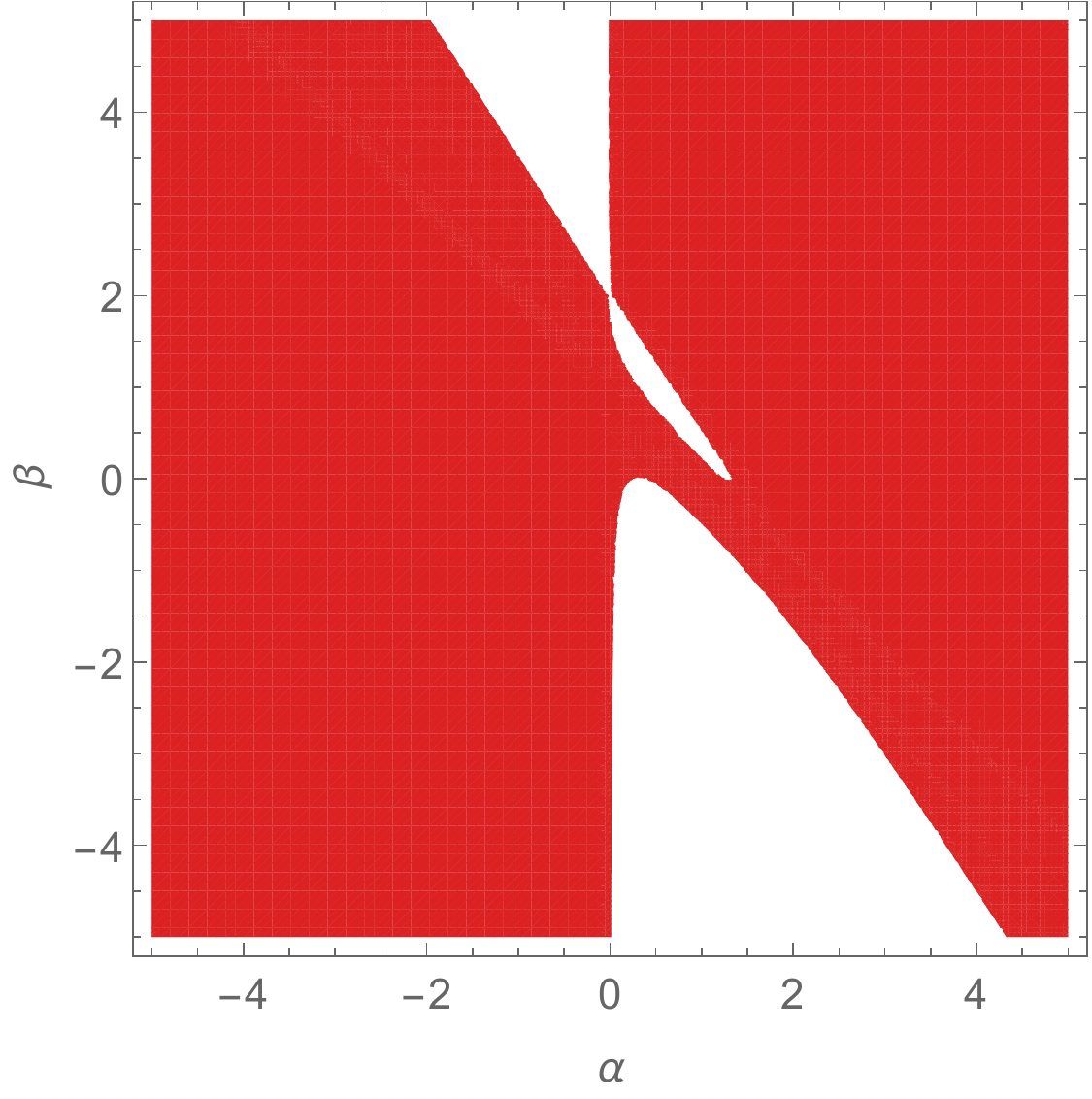}
\caption{Region of the space $\{n,m\}$ where the fixed points
$\mathcal E_{\pm}$ are defined, for the system given by
Eq.~\eqref{system1a}.}
%\vskip 1.0cm
\label{regionroot}
\end{figure}

The fixed points of the set of equations given in
Eq.~\eqref{system1a} are at most ten.
Six of them, call them $\mathcal A$,
 $\mathcal B$,
 $\mathcal C$,
 $\mathcal D$,
 $\mathcal E_+$,
 $\mathcal E_-$,
have $Y\neq0$ and are shown in Table~\ref{fixed1}.
The fixed points 
$\mathcal A$,  $\mathcal C$, and $\mathcal D$ are always unstable. 
The fixed points 
$\mathcal B$,  $\mathcal E_+$, and $\mathcal E_-$ 
can be stable or unstable depending on
the parameters $n$, $m$, and $w$.
It is very difficult to present in a compact way all the general results,
but, by inspection, one can indeed check that only the points $\mathcal B$ and $\mathcal E_{\pm}$ can be (local) attractors in the phase space, whereas the other points are always unstable. Moreover, 
point~$\mathcal B$ corresponds to a solution of the type shown in Eq.~\eqref{anasca} and therefore can lead to a singularity at finite time. Points ~$\mathcal E_{\pm}$ represent a solution approaching a constant scale factor. 
Note also that 
$\mathcal E_{\pm}$ are only defined in a specific region of the parameters $n$ and $m$ where the coordinates are real. This region is shown in Fig.~\ref{regionroot}.
The points $\mathcal B$ and $\mathcal C$
are also only defined in a specific region of the parameters
as can be worked out from Eqs.~\eqref{const1} and~\eqref{defric}.
The remaining four points out of the ten have $Y=0$ and are unstable, and 
therefore will be excluded by our analysis.
In Tables~\ref{specific1} and~\ref{specific2} one can find  an explicit analysis of the specific cases $n=1$, $m=3$, $w=1$, and  $n=-1$,
$m=3$, $w=0$.

%%%%%%%%%%%%%%%%%%%%%%%%%%%%%%%%%%%%%%%%%%%%%%%%%%%%%%%%%%%%%%%%%%%%%%%%%%%%%%%%%%%%%%%%%%%
\begin{table}[!htbp]
%Table I
\centering
\begin{tabular}{c c c c c}
\hline
Point & Coordinates & Existence & Stability & Parameter $S$ \\ \hline
\multirow{7}{*}{$\mathcal A$}
							 &$K=2n^2+2n\left(m-1\right)-1$&&&\\
                             &$X=2-n-m$&&&\\
                             &$Y=2n\left(n+m-1\right)$&&&\\
                             &$Z=\left(n+m-1\right)\left[m+3\left(n-1\right)\right]$&$3n+m\neq 3$&Saddle&$-1$\\
                             &$Q=-1$&&&\\
                             &$J=1$&&&\\
                             &$\Omega=0$&&&\\ \hline
\multirow{7}{*}{$\mathcal B$}
							 &$K=0$&&&\\
                             &$X=1$&&&\\
                             &$Y=\frac{2n}{2n+m-2}$&$n+m=2$&Saddle&\\
                             &$Z=2$&$2n+m\neq 2$&or&0\\
                             &$Q=0$&&attractor&\\
                             &$J=0$&&&\\
                             &$\Omega=0$&&&\\ \hline
\multirow{7}{*}{$\mathcal C$}
							 &$K=0$&&&\\
                             &$X=\frac{1}{n+m}$&$n+3n^3+8n^2m+7nm^2+$&&\\
                             &$Y=\frac{2\left(n+m\right)-1}{n+m}$&$+2m\left(1+m^2\right)^2-5\left(n+m\right)^2=0$&&\\
                             &$Z=\frac{1}{n+m}$&$n\neq 0$&Saddle&$\frac{-1}{\left(n+m\right)^3}$\\
                             &$Q=-\frac{1}{n+m}$&$n+m\neq 0$&&\\
                             &$J=\frac{1}{\left(n+m\right)^2}$&$n+m\neq \frac{1}{2}$&&\\
                             &$\Omega=0$&&&\\ \hline
\multirow{7}{*}{$\mathcal D$}
							 &$K=0$&&&\\
                             &$X=\frac{\left(n+m\right)\left(3w+1\right)-3\left(w+1\right)}{2\left(n+m\right)}$&$w\neq \left\{\frac{1}{3},0\right\}$&&\\
                             &$Y=\frac{4\left(n+m\right)-3\left(1+3w\right)}{n+m}$&$n+m\neq 0$&&\\
                             &$Z=\frac{\left(3w-1\right)\left[\left(n+m\right)\left(3w+1\right)-3\left(w+1\right)\right]}{4\left(n+m\right)}$&$4\left(n+m\right)\neq 3\left(1+w\right)$&Saddle&$-\frac{27}{8}\left(\frac{1+w}{n+m}\right)^3$\\
                             &$Q=-\frac{3\left(w+1\right)}{2\left(n+m\right)}$&$\left(n+m\right)\left(3w+1\right)\neq 3\left(1+w\right)$&&\\
                             &$J=\frac{9\left(w+1\right)^2}{4\left(n+m\right)^2}$&&&\\
                             &$\Omega=W\left(n,m,w\right)$&&&\\ \hline
\multirow{7}{*}{$\mathcal E_{\pm}$}
							 &$K=0$&&&\\
                             &$X=\frac{m+n\left(n+m+2\right)-\left[2\pm f\left(n,m\right)\right]}{2n\left[2\left(n+m\right)-1\right]}$&$n+m\neq 0$&&\\
                             &$Y=2-\frac{1\pm f\left(n,m\right)}{2n\left(n+m-1\right)}+\frac{3\left(1-n\right)\pm 2f\left(n,m\right)}{2n\left[2\left(n+m\right)-1\right]}$&$n+m\neq 1$&Saddle&\\
                             &$Z=g_\pm\left(n,m\right)$&$+$&or&NA\\
                             &$Q=\frac{m-\left[2\pm f\left(n,m\right)\right]+n\left[4-3\left(n+m\right)\right]}{2n\left(n+m-1\right)\left[2\left(n+m\right)-1\right]}$&additional&attractor&\\
                             &$J=h_\pm\left(n,m\right)$&conditions&&\\
                             &$\Omega=0$&&&\\ \hline
\multicolumn{5}{c}{} \\
\multicolumn{5}{c}{$W\left(n,m,w\right)=\frac{1}{2\left(3w-1\right)}\left\{\frac{8m}{n-\frac{2\left[11+3w+n\left(-4+9w+9w^2\right)\right]}{n}}-\frac{9\left(1+w\right)^2}{\left(n+m\right)^2}+\frac{3\left(1+w\right)\left[4+\left(3+9w\right)n\right]}{n\left(n+m\right)}\right\}$} \\ 
\multicolumn{5}{c}{} \\
\multicolumn{5}{c}{$f\left(n,m\right)=\sqrt{\left[2+n\left(5n-8\right)\right]^2+2m\left[\left(n-1\right)n\left(33n-38\right)-2\right]+m^2\left[1+n\left(57n-62\right)\right]+16nm^3}$} \\ 
\multicolumn{5}{c}{} \\
\multicolumn{5}{c}{$g_\pm\left(n,m\right)=\frac{1}{8n^2}\left\{1-12n+15n^2+8nm-\frac{4\left[1\pm f\left(n,m\right)\right]\left(n-1\right)}{n+m-1}+\frac{3\left(n-1\right)\left[3\left(1-n\right)\pm2f\left(n,m\right)\right]}{\left[2\left(n+m\right)-1\right]^2}+\frac{3\left(n-2\right)\left(n+1\right)\pm f\left(n,m\right)\left(n-5\right)}{2\left(n+m\right)-1}\right\}$} \\ 
\multicolumn{5}{c}{} \\
\multicolumn{5}{c}{$h_\pm\left(n,m\right)=\frac{17n^4+\left(m-2\right)^2+n^3\left(42m-52\right)\pm f\left(n,m\right)\left[2-m+n\left(3\left(n+m\right)-4\right)\right]+2n\left(m-2\right)\left[6+m\left(4m-9\right)\right]+n^2\left[56+m\left(33m-86\right)\right]}{2n^2\left(n+m-1\right)^2\left[2\left(n+m\right)-2\right]^2}$} \\ 
\multicolumn{5}{c}{} \\ \hline
\end{tabular}
\caption{Fixed points for the system given by
Eq.~\eqref{system1a}. The solution for the parameter $S$ in the fixed
point $\mathcal E_\pm$ cannot be represented in an easy way because
of its complexity. The same happens for the additional conditions
arising from the constraints~\eqref{const1} and~\eqref{defric}.}
\label{fixed1}
\end{table}
%%%%%%%%%%%%%%%%%%%%%%%%%%%%%%%%%%%%%%%%%%%%%%%%%%%%%%%%%%%%%%%%%%%%%%%%%%%%%%%%%%%%%%%%%%%%

%%%%%%%%%%%%%%%%%%%%%%%%%%%%%%%%%%%%%%%%%%%%%%%%%%%%%%%%%%%%%%%%%%%%%%%%%%%%%%%%%%%%%%%%%%
%%%%%%%%%%%%%%%%%%%%%%%%%%%%%%%%%%%%%%%%%%%%%%%%%%%%%%%%%%%%%%%%%%%%%%%%%%%%%%%%%%%%%%%%%%

%%%%%%%%%%%%%%%%%%%%%%%%%%%%%%%%%%%%%%%%%%%%%%%%%%%%%%%%%%%%%%%%%%%%%%%%%%%%%%%%%%%%%%%%%%%
\begin{table}[!htbp]
\centering
\begin{footnotesize}
\begin{tabular}{c c c c c c c c c c}
\hline
& $K$ & $X$ & $Y$ & $Z$ & $Q$ & $J$ & $\Omega$ & Stability & Parameter $S$ \\ \hline
$\mathcal A$ & 5 & $-2$ & 6 & 9 & $-1$ & 1 & 0 & Saddle & $-1$ \\ 
$\mathcal D$ & 0 & $-\frac{5}{4}$ & $\frac{5}{4}$ & $\frac{5}{4}$ & $-\frac{3}{4}$ & $\frac{9}{16}$ & $-\frac{41}{16}$ & Saddle &  $-\frac{27}{63}$ \\
$\mathcal E_-$ & 0 & $\frac{1}{14}\left(7-\sqrt{385}\right)$ & $\frac{1}{42}\left(77-\sqrt{385}\right)$ & $\frac{23}{7}-\sqrt{\frac{55}{7}}$ & $\frac{1}{42}\left(-7-\sqrt{385}\right)$ & $\frac{1}{126}\left(31+\sqrt{385}\right)$ & 0 & Saddle & $\frac{-301-19\sqrt{385}}{2646}$ \\ 
$\mathcal E_+$ & 0 & $\frac{1}{14}\left(7+\sqrt{385}\right)$ & $\frac{1}{42}\left(77+\sqrt{385}\right)$ & $\frac{23}{7}+\sqrt{\frac{55}{7}}$ & $\frac{1}{42}\left(-7+\sqrt{385}\right)$ & $\frac{1}{126}\left(31-\sqrt{385}\right)$ & 0 & attractor & $\frac{-301+19\sqrt{385}}{2646}$ \\ 
\end{tabular}
\end{footnotesize}
\caption{Fixed points for the system given by Eq.~\eqref{system1a} in the specific case $\alpha=1$, $m=3$, $w=1$.}
\vskip 1.0cm
\label{specific1}
\end{table}
%%%%%%%%%%%%%%%%%%%%%%%%%%%%%%%%%%%%%%%%%%%%%%%%%%%%%%%%%%%%%%%%%%%%%%%%%%%%%%%%%%%%%%%%%%

%%%%%%%%%%%%%%%%%%%%%%%%%%%%%%%%%%%%%%%%%%%%%%%%%%%%%%%%%%%%%%%%%%%%%%%%%%%%%%%%%%%%%%%%%%
\begin{table}[!htbp]
\centering
\begin{tabular}{c c c c c c c c c c}
\hline
& $K$ & $X$ & $Y$ & $Z$ & $Q$ & $J$ & $\Omega$ & Stability & Parameter $S$ \\ \hline
$\mathcal A$ & $-3$ & 0 & -2 & -3 & $-1$ & 1 & 0 & Saddle & $-\frac{1}{27}$ \\ 
$\mathcal B$ & 0 & 1 & 2 & 2 & 0 & 0 & 0 & attractor & 0 \\ 
$\mathcal D$ & 0 & $\frac{1}{4}$ & $\frac{5}{4}$ & $\frac{1}{8}$ & -$\frac{3}{4}$ & $\frac{9}{16}$ & $-\frac{9}{8}$ & Saddle &  $-\frac{27}{64}$ \\ 
\end{tabular}
\caption{Fixed points for the system given by Eq.~\eqref{system1a} specific case $n=-1$, $m=-3$, $w=0$.}
\vskip 1.0cm
\label{specific2}
\end{table}
%%%%%%%%%%%%%%%%%%%%%%%%%%%%%%%%%%%%%%%%%%%%%%%%%%%%%%%%%%%%%%%%%%%%%%%%%%%%%%%%%%%%%%%%%%

\newpage
\centerline{}
\newpage
%%%%%%%%%%%%%%%%%%%%%%%%%%%%%%%%%%%%%%%%%%%%%%%%%%%%%%%%%%%%%%%%%%%%%%%%%%%%%%%%%%%%%%%%%%
\subsection{The case of $\alpha R^n+\beta\mathcal R^m$ gravity}
\label{dsa2}
%%%%%%%%%%%%%%%%%%%%%%%%%%%%%%%%%%%%%%%%%%%%%%%%%%%%%%%%%%%%%%%%%%%%%%%%%%%%%%%%%%%%%%%%%%

In this section we consider that the
function $f$ has the form
$f=\alpha R^n+\beta\mathcal R^m$, for some constants $\alpha$
and $\beta$
and free exponents $n$ and $m$
which can be put in the form
$f=\alpha_* \left(\frac{R}{R_0}\right)^n+\beta_*\left(\frac{\mathcal R}{R_0}\right)^m$,
with $\alpha_*$, $\beta_*$, and $R_0$ constants,
and so the action $S^*$, say, is
$S^*=\int\sqrt{-g}\left[\alpha_* \left(\frac{R}{R_0}\right)^n+\beta_*\left(\frac{\mathcal R}{R_0}\right)^m\right] d^4x +  {S}_m^*$,
with
${S}_m^*$ the matter action.
Since multiplying the action by a constant does not affect the resultant equations of motion, we can take $\alpha_*$ out of the action and write
\be \label{act2}
S=\int\sqrt{-g}\left[\left(\frac{R}{R_0}\right)^n+\gamma_*\left(\frac{\mathcal R}{R_0}\right)^m\right] d^4x + {S}_m,
\ee
for some constant $\gamma_*=\alpha_*/\beta_*$ and defined ${S}_m=\alpha_*^{-1}{S}_m^*$.
Note that $\gamma_*$ is a parameter that allows us to select which of the two terms is dominant. For $\gamma_*\ll1$ we have a dominant $f\left(R\right)$ term and for $\gamma_*\gg1$ we have a dominant $f\left(\mathcal R\right)$ term. 
The Jacobian from Eq.~\eqref{jacobian} for this $f$ can be written in
terms of the dynamic variables and parameters as
\be
J=\frac{A^n Y^{1-n}}{18na^2H^7}.
\ee
For this Jacobian to be finite, we must exclude the value $n=0$ from the analysis and also constrain our results for the fixed points to have values for the variables $Y$ and $A$ different from zero.

The dynamical functions in Eq.~\eqref{GenFunct} become
\beq\label{GenFunctb}
&&\textbf{A}=\frac{\gamma_* mZ^{m-1}A^{n-m}}{ n Y^{n-1}},\ \ \ \textbf{B}=\frac{Y}{n}+\frac{\gamma_* Z^mA^{n-m}}{n Y^{n-1}},\ \ \ \textbf{D}=\frac{2Z}{m-1},\ \ \ \textbf{G}=\frac{m-2}{2Z},\nonumber \\
&&\textbf{H}=\frac{n \left(n-1\right)\left(n-2\right)Y^{n-3}A^{m-n}}{2\gamma_* m \left(m-1\right)Z^{n-2}},\ \ \ \textbf{I}=\frac{n \left(n-1\right)Y^{n-2}A^{m-n}}{\gamma_* m \left(m-1\right)Z^{m-2}},\ \ \ \textbf{C}=\textbf{E}=\textbf{F}=0,
\eeq
and, using the constraints from Eqs.~\eqref{defric} and~\eqref{constrZ}, the dynamical system from Eq.~\eqref{simpsys} becomes
\beq
K'&=&2K\left(K-Y+1\right),\nonumber \\
X'&=&Z-X\left(X+Y-1\right)+K\left(X-1\right),\nonumber \\
Y'&=&Y\left\{2\left(2+K-Y\right)+\frac{1}{1-n}\left\{1+\frac{n\left(n-1\right)Y^{n-2}Z^{2-m}}{m\left(m-1\right)\gamma_* A^{n-m}}+K-Y+\frac{m\gamma_* Y^{1-n}Z^{m-1}}{n A^{m-n}}\left[K+X^2-Z\right]-\Omega\right\}\right\}, \label{system1b}\\
Z'&=&2Z\left(\frac{X-1}{m-1}+2+K-Y\right),\nonumber\\
\Omega'&=&-\Omega\left[-2+3w-\frac{Y}{n}\left(1+\frac{\gamma_* Z^m}{Y^n A^{m-n}}\right)-3\left(K-Y\right)-\frac{\gamma_* m Z^{m-1}}{n Y^{n-1} A^{m-n}}\left(K+X^2-Z\right)+\Omega\right]\nonumber \\
A'&=&2A\left(2+K-Y\right).\nonumber
\eeq
The system of equations in Eq.~\eqref{system1b}  presents divergences for specific values of the parameters  for $m=0, m=1,$ and $n=1$, which implies that our formulation is not valid for these cases. Indeed, when this is the case the functions in Eq.~\eqref{GenFunctb} are divergent and the analysis should be performed starting again from the cosmological Eqs.~\eqref{dinF} and~\eqref{dinR}. The dynamical system also presents some divergences for  $Y=0$ and $Z=0$ which are due to the  very structure of the  gravitation field equations for this choice of the action. Because of these singularities the dynamical system is not $C(1)$ in the entire phase space and one can use the standard analysis tool of the phase space only when $Y, Z\neq0$. We will pursue this kind of analysis here.

The system given in Eq.~\eqref{system1b} presents the  $K=0$ and $\Omega=0$ invariant submanifolds together with the invariant submanifolds $Z=0$ and $A=0$. The presence of the latter submanifolds allows us to solve partially the problem about the singularities in the phase space. The presence of the $Z=0$  submanifold implies once again that no orbit will cross this surface. However the issue remains for the $Y=0$ hypersurface. The presence of this submanifold would prevent the existence of a global attractor for this case but, as we will see, this model does not have any finite attractors. Knowing this, it is possible again to analyze discrete sets of initial conditions and parameters and verify the time-asymptotic state for the system.

Much in the same way of the $f(R)$-gravity for this model, the Jacobian $J$ vanishes for all but one of the eighteen fixed points.  Since one can prove that all of these fixed points correspond to singular states of the field equations and that they are unstable, we will ignore them. Therefore the theory has only one relevant fixed point which we will call $\mathcal B$.  As from Table~\ref{fixed2a} the existence of  $\mathcal B$ depends on the values of the parameters $n$, $m$ and $\gamma_*$ and the point is always a saddle point. If $n\neq m$ it only exists if $m-n$ is an odd number. For $m-n=0$, the equation for the variable $A$ decouples from the rest of the system. However, the equation for $A$ can be considered as an extra constraint for the system which carries a memory of the properties of the complete system, such as divergences for $A=0$ and specific values of $\gamma_*$ that allow the existence of point $\mathcal B$. More specifically, if $m=n=2$, $\mathcal B$ exists for any value of $\gamma_*$, whereas for $m=n\neq 2$ then $\mathcal B$ exists only for $\gamma_*=-1$.  

For any value of the parameters, $\mathcal B$ is associated to $S=0$, and therefore it corresponds to a solution of the type shown in Eq.~\eqref{anasca}. This implies that the theory can incur in a singularity at finite time.

%%%%%%%%%%%%%%%%%%%%%%%%%%%%%%%%%%%%%%%%%%%%%%%%%%%%%%%%%%%%%%%%%%%%%%%%%%%%%%%%%%%%%%%%%%%%
\begin{table}
\centering
\begin{tabular}{c c c c c}
\hline
Set & Coordinates & Existence & Stability & Parameter $S$ \\ \hline
\multirow{9}{*}{$\mathcal B$}
							 &$K=0$&&&\\
                             &$X=1$&&&\\
                             &$Y=2$&$n\neq\{0,1\}$&&\\
                             &$Z=2$&&&\\
                             &$Q=0$&If $n=m\neq 2$, then $\gamma_*=-1$&Saddle&0\\
                             &$J=0$&&&\\
                             &$\Omega=0$&If $n\neq m$, then $n-m=$ odd&&\\
                             &$A\neq 0$, if $n=m$&&&\\ 
                             &$A=-2\left(\frac{n-2}{\gamma_*\left(m-2\right)}\right)^\frac{1}{n-m}$, if $n\neq m$&&&\\ \hline
\end{tabular}
\caption{Fixed points for the system given by Eq.~\eqref{system1b}.}
\label{fixed2a}
\end{table}
%%%%%%%%%%%%%%%%%%%%%%%%%%%%%%%%%%%%%%%%%%%%%%%%%%%%%%%%%%%%%%%%%%%%%%%%%%%%%%%%%%%%%%%%%%%%

%\newpage
%\centerline{}
\newpage

%%%%%%%%%%%%%%%%%%%%%%%%%%%%%%%%%%%%%%%%%%%%%%%%%%%%%%%%%%%%%%%%%%%%%%%%%%%%%%%%%%%%%%%%%%%%
\subsection{The case of $\exp\left(\frac{R}{\mathcal R}\right)$ gravity}
%%%%%%%%%%%%%%%%%%%%%%%%%%%%%%%%%%%%%%%%%%%%%%%%%%%%%%%%%%%%%%%%%%%%%%%%%%%%%%%%%%%%%%%%%%%%

In this section we consider that the
function $f$ has the form
$f=\alpha_* \exp\left(\frac{R}{\mathcal R}\right)$, for some constant $\alpha_*$,
and so the action $S^*$, say, is
$S^*=\int\sqrt{-g}\,\alpha_* \exp\left(\frac{R}{\mathcal R}\right) d^4x +  {S}_m^*$,
with
${S}_m^*$ the matter action.
Since multiplying the action by a constant does not affect the resultant equations of motion, we can take $\alpha_*$ out of the action and write
\be\label{act3}
S=\int\sqrt{-g}\,\exp\left(\frac{R}{\mathcal R}\right) d^4x + {S}_m,
\ee
where ${S}_m=\alpha_*^{-1}{S}_m^*$ is the dimensionless matter action.
The motivation to test a model of the form of Eq.~\eqref{act3} is that
exponential functions are quite general and lead to interesting results.
The Jacobian from Eq.~\eqref{jacobian} for this case can be written in terms of the dynamic variables and parameters as
\be
J=\frac{e^{\frac{Y}{Z}}Z}{18a^2H^7}.
\ee
For this Jacobian to be finite, we must constrain our results for the fixed points to have values for the variable $Z$ different from zero.
The dynamical functions in Eq.~\eqref{GenFunct} become in this case,
\beq
&&\textbf{A}=-\frac{Y}{Z},\ \ \ \textbf{B}=Z,\ \ \ \textbf{C}=-\frac{Z\left(Y+Z\right)}{Y\left(Y+2Z\right)},\ \ \ \textbf{D}=-\frac{2Z^2}{Y+2Z},\ \ \ \textbf{E}=\frac{Y^2+4YZ+2Z^2}{2Y^2Z+4YZ^2},\nonumber \\
&&\textbf{F}=-\frac{1}{2Y},\ \ \ \textbf{G}=-\frac{Y^2+6YZ+6Z^2}{2Z^2\left(Y+2Z\right)},\ \ \ \textbf{H}=\frac{Z}{2Y^2+4YZ},\ \ \ \textbf{I}=\frac{Z^2}{Y^2+2YZ},
\eeq
and the dynamical system from Eq.~\eqref{simpsys} becomes
\beq
K'&=&2K\left(K-Y+1\right),\nonumber \\
X'&=&Z-X\left(X+Y-1\right)+K\left(X-1\right),\nonumber \\
Y'&=&Y\left\{2\left(2+K-Y\right)+\left(2+\frac{Y}{Z}\right)\left\{1+Z+K-Y-\frac{Y}{Z}\left[K-\frac{2Z\left(Y+Z\right)}{Y\left(Y+2Z\right)}\left(X-1\right)+X^2-Z\right]-\Omega\right\}\right\},\label{system1c} \\
Z'&=&\frac{2\left(Y+Z\right)^2}{Z^2}\left(2+K-Y\right)-\nonumber \\
&&-\frac{2Y}{Z}\left(Y+2Z\right)\left[-\frac{Z\left(X-1\right)}{Y+2Z}+2+K-Y\right]+\left(Y+Z\right)\left[1+Z+K-Y-\Omega-\frac{Y}{Z}\left(K+X^2-Z\right)\right],\nonumber\\
\Omega'&=&-\Omega\left[-2+3w-Z-3\left(K-Y\right)+\frac{Y}{Z}\left(K+X^2-Z\right)+\Omega\right]\nonumber \\
A'&=&2A\left(2+K-Y\right).\nonumber
\eeq
where we have used the constraints in Eqs.~\eqref{defric} and~\eqref{constrZ}.

The system of equations in Eq.~\eqref{system1c} has divergences for specific values of $Y$ and $Z$.  These divergences occur for $Y=0$, $Z=0$, and $Y+2Z=0$, and are due to the  very structure of the  gravitation field equations for this choice of the action. Because of these singularities the dynamical system is not $C(1)$ in the entire phase space and one can use the standard analysis tool of the phase space only when $Y, Z\neq0$ and $Y\neq -2Z$. We will pursue this kind of analysis here.
The system also presents the usual $K=0$ and $\Omega=0$ invariant submanifolds together with the invariant submanifold $Z=0$. The presence of the latter submanifolds allows us to solve partially the problem about the singularities in the phase space. The presence of the $Z=0$ submanifold implies again that no orbit will cross this surface. However the issue remains for the $Y=0$ hypersurface. The presence of global attractors is also prevented in this case due to the existence of this submanifold. Such attractor should have  $Z=0$, $K=0$ and $\Omega=0$, which would correspond to a singular state for the theory. We can once again use this information to discriminate sets of initial conditions and of parameters values to analyze the time-asymptotic state for the system.

The system given by Eq.~\eqref{system1c} presents at most three fixed points, which are shown in Table~\ref{fixed3} with their stability and associated solution.

%%%%%%%%%%%%%%%%%%%%%%%%%%%%%%%%%%%%%%%%%%%%%%%%%%%%%%%%%%%%%%%%%%%%%%%%%%%%%%%%%%%%%%%%%%%%
\begin{table}[!h]
\centering
\begin{tabular}{c c c c c}
\hline
Point & Coordinates & Stability & Parameter $S$ \\ \hline
\multirow{8}{*}{$\mathcal A$}
							 &$K=-6$&&\\
                             &$X=2$&&\\
                             &$Y=-5$&&\\
                             &$Z=-2$&Saddle&$-1$\\
                             &$Q=-1$&&\\
                             &$J=1$&&\\
                             &$\Omega=0$&&\\ \hline
\multirow{8}{*}{$\mathcal E_{\pm}$}
							 &$K=0$&&\\
                             &$X=-\frac{1}{2}\left(5\pm \sqrt{33}\right)$&&\\
                             &$Y=\frac{1}{2}\left(11\pm\sqrt{33}\right)$&$\mathcal E_+$: Saddle&\\
                             &$Z=-\left(5\pm\sqrt{33}\right)$&&$\frac{1}{2}\left(259\pm 45\sqrt{33}\right)$\\
                             &$Q=\frac{1}{2}\left(7\pm\sqrt{33}\right)$&$\mathcal E_-$: Attractor&\\
                             &$J=\frac{1}{2}\left(41\pm\sqrt{33}\right)$&&\\
                             &$\Omega=0$&&\\ \hline
\end{tabular}
\caption{Fixed points for the system given by Eq.~\eqref{system1c}.}
\label{fixed3}
\end{table}
%%%%%%%%%%%%%%%%%%%%%%%%%%%%%%%%%%%%%%%%%%%%%%%%%%%%%%%%%%%%%%%%%%%%%%%%%%%%%%%%%%%%%%%%%%%%

%%%%%%%%%%%%%%%%%%%%%%%%%%%%%%%%%%%%%%%%%%%%%%%%%%%%%%%%%%%%%%%%%%%%%%%%%%%%%%%%%%%%%%%%%%%%
\subsection{The case of $R\exp\left(\frac{\mathcal  R}{R}\right)$ gravity}\label{FunLink}
%%%%%%%%%%%%%%%%%%%%%%%%%%%%%%%%%%%%%%%%%%%%%%%%%%%%%%%%%%%%%%%%%%%%%%%%%%%%%%%%%%%%%%%%%%%%

In this section we consider that the
function $f$ has the form
$f=\alpha_*\,R\, \exp\left(\frac{\mathcal R}{ R}\right)$, for some constant $\alpha_*$,
and so the action $S^*$, say, is
$S^*=\int\sqrt{-g}\,\alpha_* \,R\,\exp\left(\frac{\mathcal R}{ R}\right) d^4x +  {S}_m^*$,
with
${S}_m^*$ the matter action.
Since multiplying the action by a constant does not affect the resultant equations of motion, we can take $\alpha_*$ out of the action and write
\be\label{act4}
S=\int\sqrt{-g}\,R\,\exp\left(\frac{\mathcal R}{ R}\right) d^4x + {S}_m,
\ee
where ${S}_m=\alpha_*^{-1}{S}_m^*$ is the dimensionless matter action.
This particular form of $f$ satisfies Eq.~\eqref{Cond2Ord}, and therefore the field equations are effectively of order 2.
 
In terms of Eq.~\eqref{GenFunct}, the set of equations~\eqref{Cond2Ord}  read
\begin{align}
& \textbf{C}^2-\textbf{I}=0,\\
& \textbf{F}+\textbf{G}\textbf{C}^2-2\textbf{E}\textbf{C}=0,\\
& \textbf{H}-3\textbf{C}\textbf{F}+3\textbf{C}^2\textbf{E}-\textbf{C}^3\textbf{G}=0,
\end{align}
and the cosmological equations, Eqs.~\eqref{dinF} and~\eqref{dinR}, can be written
as
\be\label{friedsimple}
\frac{1}{Y-Z}\left[Y\left(1+2K+X^2-\Omega\right)+Z\left(1-Z-2X+\Omega\right)\right]=0,
\ee
\be\label{raichsimple}
\frac{1}{Y-Z}\left\{Y\left[2\left(Q-K+Z-1\right)+2X\left(4-3X\right)+\left(1+3w\right)\Omega\right]-Z\left[2\left(2-K+Z\right)+2X\left(X-4\right)+\left(1+3w\right)\right]\right\}=0\,,
\ee
respectively.
At this point, using Eq.~\eqref{defric}, we can write $Y$, $Z$ in terms of $X, K, \Omega$ and
substituting in the of equations given in Eq.~\eqref{SysGen} we obtain
\beq
K'&=&-2K\left(1+Q\right),\nonumber \\
X'&=&-K-\left(1+Q\right)X-X^2+Z,\label{systemd}\\
\Omega'&=&\frac{\Omega}{Y-Z}\left[Z\left(1+3w+2Q+2X\right)-Y\left(3+3w+2Q\right)\right],\nonumber
\eeq
where $Y=Y(X, K, \Omega)$, $Z=Z(X, K, \Omega)$ and $Q=Q(X, K, \Omega)$ have not been fully substituted for the sake of simplicity.  
The Jacobian in Eq.~\eqref{jacobian} for this case can be written in terms of the dynamic variables and parameters as
\be
J=\frac{Ye^{-\frac{Z}{Y}}}{108a^2H^9\left(Y-Z\right)}.
\ee
For this Jacobian to be regular, we must exclude the fixed points that have values of $Y=Z$ or $Y=0$, which also represent divergences for the system in Eq.~\eqref{systemd} and  the very gravitation field equations for this choice of the action. Because of these singularities the dynamical system given in Eqs.~\eqref{systemd} is not $C(1)$ in the entire phase space, and one can use the standard analysis tool of the phase space only when $Y\neq Z$ and $Z\neq 0$. We will pursue this kind of analysis here.
The system given in Eqs.~\eqref{systemd} also presents the usual $K=0$ and $\Omega=0$ invariant submanifolds together with the invariant submanifold $Z=0$.  The presence of this last submanifold allows us to solve partially the problem about the singularities in the phase space. The analysis is the same as before, i.e., the presence of the $Z=0$  submanifold implies that no orbit will cross this surface, which also prevents the presence of a global attractor  for this case. Such attractor should have  $Z=0$, $K=0$, and $\Omega=0$ and therefore would correspond to a singular state for the theory. We can therefore discriminate sets of initial conditions and of parameters values which give rise to a given time-asymptotic state for the system. 

The system given in Eq.~\eqref{systemd} presents at most three fixed points, which are shown in Table~\ref{fixed4} with their stability and associated solution. These points are all nonhyperbolic, i.e., linear analysis cannot be used to ascertain their stability. A standard tool for the analysis for this type of points is the analysis of the central manifold \cite{Wiggins}. The method consists in rewriting the system given in
Eq.~\eqref{systemd} in terms of new variables $(U_1,  U_2, U_3)$ in the form,
\beq
U_1'=AU_1+F_1\left(U_1,U_2,U_3\right),\nonumber \\
U_2'=BU_2+F_2\left(U_1,U_2,U_3\right),\\
U_3'=CU_3+F_3\left(U_1,U_2,U_3\right),\nonumber \\
\eeq
where $A$, $B$, and $C$ are constants and the functions $F_i$, with $\{i,j\}=1,2,3$,
respect the conditions $F_i\left(0,0,0\right)=0$ and $\frac{\partial F_i}{\partial U_j}\left(0,0,0\right)=0$. Supposing that the quantity $A$ has zero real part, the variables $U_2$ and $U_3$ can be written as
\beq
U_2=h_2\left(U_1\right),\nonumber \\
U_3=h_3\left(U_1\right),
\eeq
and the center manifold can be defined by the equations,
\beq
h_2'\left(U_1\right)\left[A U_1+F_1\left(U_1,h_2\left(U_1\right),h_3\left(U_1\right)\right)\right]-Bh_2\left(U_1\right)-F_2\left(U_1,h_2\left(U_1\right),h_3\left(U_1\right)\right)=0\,,,\nonumber \\
h_3'\left(U_1\right)\left[A U_1+F_1\left(U_1,h_2\left(U_1\right),h_3\left(U_1\right)\right)\right]-Ch_3\left(U_1\right)-F_3\left(U_1,h_2\left(U_1\right),h_3\left(U_1\right)\right)=0\,,
\eeq
which can be solved by series. The stability of the nonhyperbolic point will be then determined by the structure of the equation,
\beq\label{CMeqGen}
U_1'=AU_1+F_1\left(U_1,h_2(U_1),h_3(U_1)\right).
\eeq
For point $\mathcal A$, the variable transformation is 
\beq
U_1=K,\\
U_2=X-\frac{1}{2},\\
U_3=\Omega,
\eeq
and Eq.~\eqref{CMeqGen} takes the form,
\be
U_1'=\frac{8}{5}U_1^2+\mathcal O\left(U_1^3\right),
\ee
which implies that this point is a saddle. For point $\mathcal B$, instead, the variable transformation is 
\beq
U_1=K+1,\\
U_2=X-\frac{1}{2}\left(1-3w\right),\\
U_3=\Omega+1+3w,
\eeq
and Eq.~\eqref{CMeqGen} takes the form
\be
U_1'=2U_1^2+\mathcal O\left(U_1^3\right),
\ee
which implies again that this point is a saddle. Finally, for point $\mathcal C$ the variable transformation is 
\beq
U_1=K.\\
U_2=X+\frac{3}{2}\left(w-1\right),\\
U_3=\Omega-2+3w,
\eeq
and Eq.~\eqref{CMeqGen} takes the form,
\be
U_1'=2\left[\frac{1}{3\left(1-w\right)}+\frac{1}{9w-5}\right]U_1^2+\mathcal O\left(U_1^3\right).
\ee
For $0<w<1$ also this point is a saddle.

The solutions associated to the fixed points can be found by the relation,
\beq
\frac{\dot{H}}{H^2}=Q\,,
\eeq
and using $Q=Q(X, K, \Omega)$ obtained by Eqs.~\eqref{friedsimple} and~\eqref{raichsimple} and evaluated at the fixed point. In general we have  
\beq
a\left(t\right)&=&a_0\exp\left(H_0t\right),\ \ \ \ \ {\mathcal q}=0,\nonumber \\
a\left(t\right)&=&a_0\left(t-t_0\right)^{-\frac{1}{{\mathcal q}}},\ \ \ \ \ {\mathcal q}\neq 0\label{solad},
\eeq
where $H_0$, $a_0$ and $t_0$ are constants of integration and ${\mathcal q}$ is the value of $Q$ at the fixed point. Notice that the fixed points are characterized by only two different values of ${\mathcal q}$, i.e., $-1$ and $-2$. Using Eq.~\eqref{solad}, we verify that the solution for ${\mathcal q}=-1$ corresponds to a linearly growing scale factor, whereas the solution for ${\mathcal q}=-2$ corresponds to a solution for the scale factor that grows with $\sqrt{t}$.

%%%%%%%%%%%%%%%%%%%%%%%%%%%%%%%%%%%%%%%%%%%%%%%%%%%%%%%%%%%%%%%%%%%%%%%%%%%%%%%%%%%%%%%%%%
\begin{table}[!h]
\centering
\begin{tabular}{c c c c c}
\hline
Point & Coordinates & Stability & Parameter Q \\ \hline
\multirow{6}{*}{$\mathcal A$}
							 &$K=0$&&\\
                             &$X=\frac{1}{2}$&&\\
                             &$Y=0$&Saddle&$-2$\\
                             &$Z=-\frac{1}{4}$&&\\
                             &$\Omega=0$&&\\ \hline
\multirow{6}{*}{$\mathcal B$}
							 &$K=-1$&&\\
                             &$X=\frac{1}{2}\left(1-3w\right)$&&\\
                             &$Y=0$&Saddle&$-1$\\
                             &$Z=\frac{3}{4}\left(1+3w\right)\left(w-1\right)$&&\\
                             &$\Omega=-\left(3w+1\right)$&&\\ \hline
\multirow{6}{*}{$\mathcal C$}
							 &$K=0$&&\\
                             &$X=-\frac{3}{2}\left(w-1\right)$&&\\
                             &$Y=0$&Saddle&$-2$\\
                             &$Z=\frac{3}{4}\left(3w-1\right)\left(w-1\right)$&&\\
                             &$\Omega=2-3w$&&\\ \hline
\end{tabular}
\caption{Fixed points for the system given by Eq.~\eqref{systemd}.}
\label{fixed4}
\end{table}
%%%%%%%%%%%%%%%%%%%%%%%%%%%%%%%%%%%%%%%%%%%%%%%%%%%%%%%%%%%%%%%%%%%%%%%%%%%%%%%%%%%%%%%%%%%

One of the motivations behind the choice to analyze an action of the
form of Eq.~\eqref{act4} was the comparison with the results recently
obtained in \cite{Rosa:2017jld}. Indeed, in \cite{Rosa:2017jld} some
forms of the function $f\left(R,\mathcal R\right)$, including the one
of Eq.~\eqref{act4}, were obtained by reconstruction from a given
cosmological solution. The phase space analysis we have performed
allows us to understand the stability of such solutions, which was
impossible to obtain by the reconstruction method of
\cite{Rosa:2017jld}. In particular, point $\mathcal A$ corresponds to
the solution found in \cite{Rosa:2017jld}, i.e., a flat $K=0$, vacuum
$\Omega=0$, universe with $a\left(t\right)$ proportional to $\sqrt{t}$
and we determined here that such solution is unstable. This shows that
we can use the phase space to determine the stability of the
solutions obtained in our previous work even if these results were
obtained using a nontrivial redefinition of the action. It should be
stressed, however, that it is not necessarily true that an exact
solution found for the cosmological equation of a given theory
corresponds to a fixed point of our phase space.  For example, in
\cite{Rosa:2017jld}, an exact nonflat vacuum solution was found  for
a theory with an action of the form of Eq.~\eqref{act4}
which does not correspond
to a fixed point of the phase space. However, in general phase space
analysis it is useful to understand in a deeper way not only the
stability of the solution, but also the consequences of the
reorganization of the degrees of freedom that is often employed to
analyze this class of theories.

\section{Connection with observational data}
\label{data}

\subsection{The fourth-order model Eq.~\eqref{act1}}

The analysis above might appear of mathematical
interest only and
of course 
relevant physical information is required.  We now show that,
although there are some intrinsic difficulties, the results above can
be used to deduce nontrivial features of the cosmologies of hybrid
metric-Palatini theories.  Here we select a suitable model
from the previous section, the one of
order 4, and deduce by numerical
integration of the dynamical system equations the behavior of the
cosmological parameters consistent with a set of initial condition
consistent with observations.

Thus, let us analyze a particular form of the action given in
Eq.~\eqref{act1} and the system of equations given in
Eq.~\eqref{system1a} with $n=m=2$. Furthermore, we consider the
matter distribution to be dust, i.e., $w=0$, as in cosmology galaxies
are often considered to be the matter elements and they do not
interact with each other apart from their gravitational
interaction. In this particular case, the dynamical system
given in Eq.~\eqref{system1a} is simplified to
\beq
K'&=&2K\left(1+K-Y\right),\nonumber\\
X'&=&\left(1+K\right)X+Z-K-X\left(X+Y\right),\nonumber\\
Y'&=&\frac{Y}{6}\left[\frac{Y\left(2K+2X^2-15Z\right)}{Z}+2\left(9+7K+4X-\Omega\right)\right],\label{systemint}\\
Z'&=&\frac{1}{3}\left[-2X^2Y-2K\left(Y-2Z\right)-2XZ+Z\left(12-3Y+2\Omega\right)\right],\nonumber \\
\Omega'&=&\frac{\Omega}{2Z}\left[2X^2Y+2K\left(Y+3Z\right)+Z\left(4-7Y-2\Omega\right)\right].\nonumber
\eeq

\begin{figure}[h]
\centering
\includegraphics[scale=0.7]{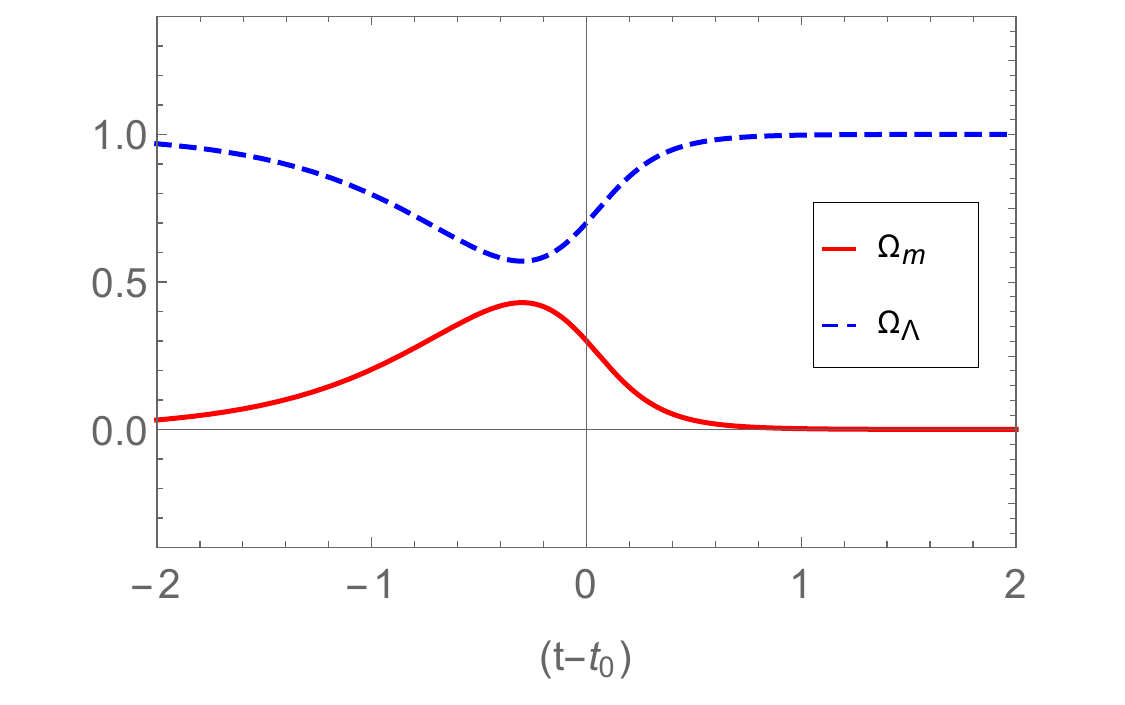}
\includegraphics[scale=0.67]{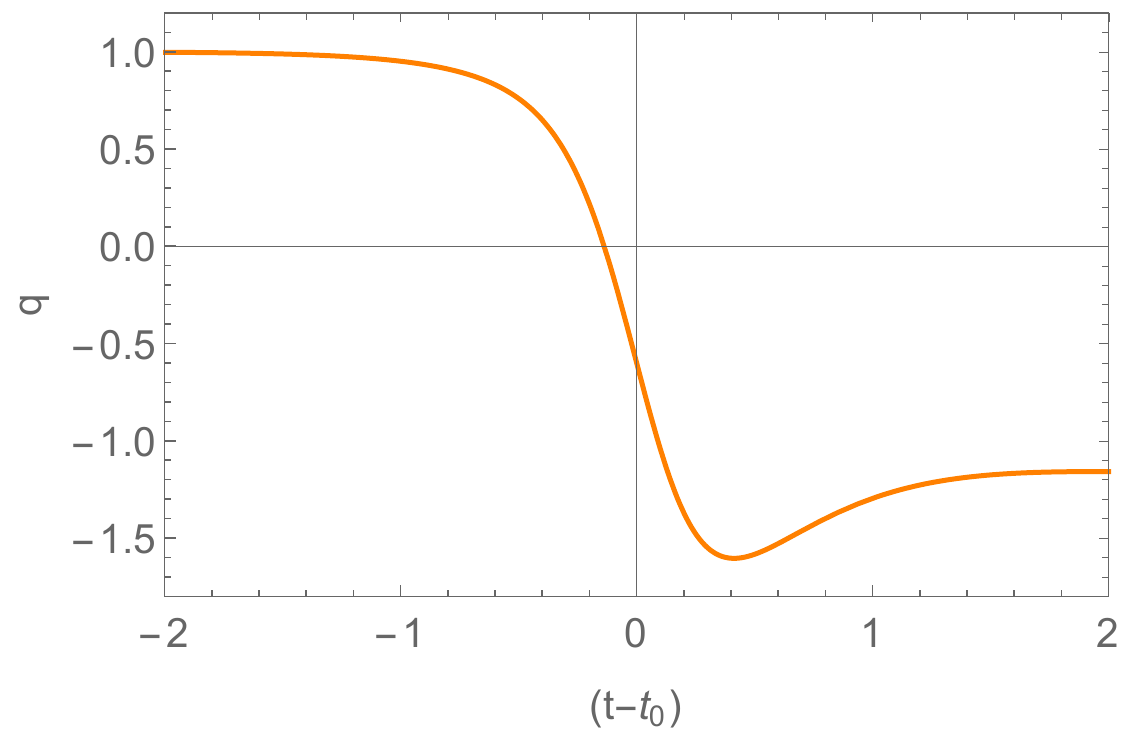}
\caption{Results for the numerical integration of the system in
Eq.~\eqref{systemint}, a simplified version of Eq.~\eqref{system1a},
with $w=0$ and under initial conditions consistent with the observational
constraints. The parameters
$\Omega_m$, $\Omega_\Lambda$, and $q$ are plotted as functions of time.
}
%\vskip 1.0cm
\label{fig:numint}
\end{figure}

To find the initial conditions
$\left\{K_0,X_0,Y_0,Z_0,Q_0,J_0,\Omega_0\right\}$ for the integration,
we proceed as follows.
The spacial curvature of the Universe has been
measured and it is approximately zero, i.e., the Universe seems to be
flat, and thus we assume $K_0=0$
\cite{Aghanim:2018eyx}.
Let us take the value for ${q{}}$ the observationally measured value,
i.e., ${q{}}=-0.6$.
Comparing
Eq.~\eqref{qdec0} with
Eq.~\eqref{qpar}, we verify that the
assumed value for ${q{}}$ imply
the value $Q_0=-0.4$
For the parameter ${j{}}$, we note that it as
not been measure to date, although the value ${j{}}=1$ 
 is consistent
with a few models explored recently \cite{Mamon:2018dxf}, and so we
take this value.
Comparing
Eq.~\eqref{jerk0} with~\eqref{jpar}, we verify that the
assumed values for  ${j{}}$ imply
the value
$J_0=4.8$. Now, from Eq.~\eqref{defric} we can compute
that $Y_0$ has the value
$Y_0=1.6$.
Equation~\eqref{const1} can be put in the form
$1+K=\Omega+\Omega_\Lambda$, where
$\Omega_\Lambda\equiv4\left(1-X\right)+\frac{3}{Y}\left[J-2K+Q
\left(4+Q\right))\right]+\frac{3Y}{2}-\frac{Y}{Z}\left(K+X^2\right)$
is the dark energy density.
The values of the parameters $\Omega$ and $\Omega_\Lambda$
have been measured and they are roughly $\Omega_0=0.3$ and
$\Omega_{\Lambda_0}=0.7$. Under these assumptions
the latter equation defining $\Omega_\Lambda$
becomes an equation for $Z$ as a function of $X$ as
$Z=\frac{4X^2}{30-10X}$.
As there are no other constraints in the problem, $Z$ and $X$
do not have a unique value consistent with the observations. Much in the
same way of what it is done in the case of scalar field cosmologies we
will set a specific, but arbitrary, value for either $X_0$ or $Z_0$
and obtain the other variable via the above relation. 
A limitation inherent of the approach that we have used to
construct the dynamical system equations is that the phase space is
not compact. Such choice was forced by the high generality of the
classes of theories we considered. In terms of the numerical
integration this issue induces the appearance of spurious divergences
in the numerical integration of the dynamical equations.  We thus have
to carefully select a pair $\left\{X_0,Z_0\right\}$ for which the
orbit in the phase space stays regular. A combination that preserves
this regularity is
$\left\{X_0,Z_0\right\}=\left\{2.7,9.72\right\}$. This completes our
set of initial conditions and we are now ready to proceed to the
numerical integration of the system in Eq.~\eqref{systemint}. The
results for the numerical integration for the matter density $\Omega$,
the dark energy density $\Omega_\Lambda$, and the deceleration
parameter ${q{}}$ are plotted in Fig.~\ref{fig:numint}.
%%%%%%%%%%%%%%%%%%%%%%%%%%%%%%%%%%%%%%%%%%%%%%%%%%%%%%%%%%%%%
%%%%%%%%%%%%%%%%%%%%%%%%%%%%%%%%%%%%%%%%%%%%%%%%%%%%%%%

%\newpage
It is clear that this model is consistent with the late-time
cosmic acceleration as ${q{}}$ becomes eventually negative, and stays
negative, just before the present era. Notice that the dark energy
component is always dominating over the matter component, but this
does not imply ${q{}}<0$ at all time. This implies that
$\Omega_\Lambda$ does not always act as a dark component. It would
be interesting to understand how this features influences the the
formation of large scale structures. Another consequence of the early
${q{}}>0$ phase is that this model cannot describe early time
inflation, suggesting that this type of model might need some form of
early time correction. This is interesting as it appears that the
Palatini terms are able to neutralize the effect of the higher order
terms, which notoriously produce inflationary behavior at early time.

One could try other ways to compare
our cosmological model to observations. In the case of the
supernovae Ia, for example, it is possible to make a statistical
analysis of the distance modulus in order to constrain the
parameter of the theory, for an example in another context see
\cite{Carloni:2018ioq}. We will not attempt such analysis here for,
essentially, two reasons. First because our aim in the following
will be only to show that there is a clear connection between our
very general phase space analysis and observations. Second, such
analysis is not necessarily immediate. The statistical analysis
mentioned above, for example, would require finding an expression of
$H$ from the Friedmann equation, Eq.~\eqref{dinF}, which is a nonlinear
equation in $H$.

\subsection{The second-order model Eq.~\eqref{act4}}

In order to proceed extracting interesting physical information
on the  mathematical analysis of the previous section,
we select another suitable
interesting model, the one of second-order.
Again, we are committed to
find nontrivial results
from cosmologies provided by hybrid
metric-Palatini theories
and work out through numerics on the 
dynamical system equations the behavior of the cosmological parameters
consistent with a set of
initial condition consistent with the  observations.

\begin{figure}[h]
\centering
\includegraphics[scale=0.7]{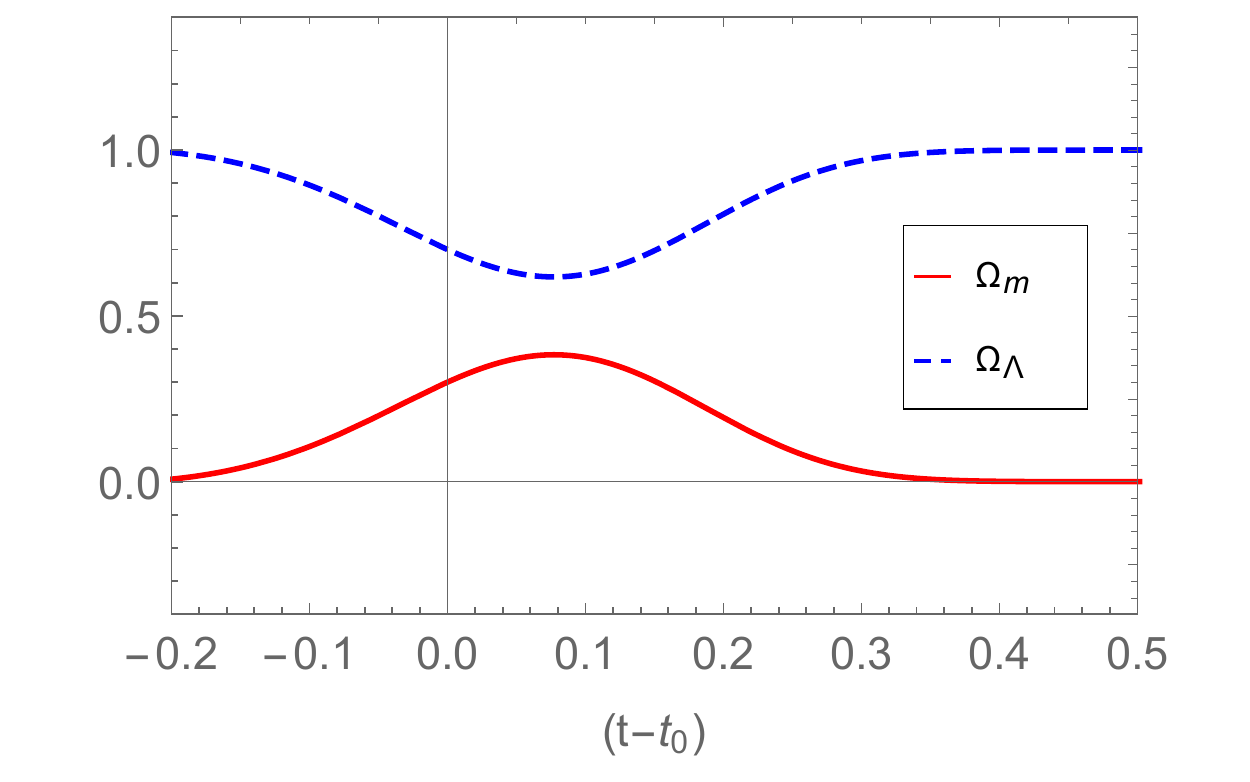}
\includegraphics[scale=0.67]{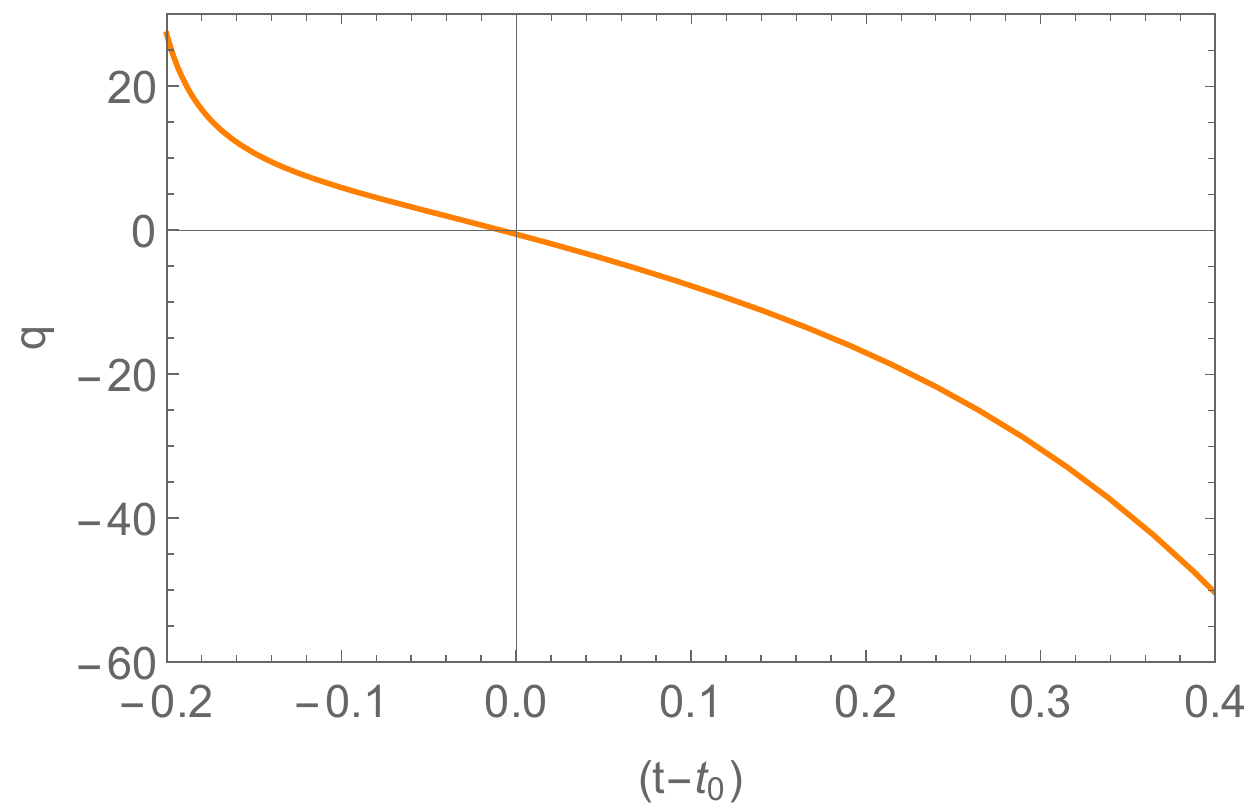}
\caption{Results for the numerical integration of the system in
Eq.~\eqref{systemd} with $w=0$ and under initial conditions consistent
with the observational constraints. The parameters
$\Omega_m$, $\Omega_\Lambda$, and $q$ are plotted as functions of time.}
%\vskip 1.0cm
\label{fig:numint2}
\end{figure}

Let us now consider the action given in Eq.~\eqref{act4}. Again, we
shall consider $w=0$, and the dynamical system will be the one given
in Eq.~\eqref{systemd}. As can be seen from Eq.~\eqref{raichsimple},
in this case the Raychaudhuri equation does not depend on the snap
parameter $s$ anymore and thus it can be used as an extra constraint
to cancel the indetermination between $X$ and $Z$ from the previous
model. In this way we have $X_0\sim -5.06$ and $Z_0\sim 3.68$. We can
now proceed to the numerical integration, whose results for $\Omega$,
$\Omega_\Lambda$ and ${q{}}$ are plotted in Fig.~\ref{fig:numint2}.

%%%%%%%%%%%%%%%%%%%%%%%%%%%%%%%%%%%%%%%%%%%%%%%%%%%%%%%%%%%%%%%%%%%%%%%%%%%%%%%%%%%%%%%%%%
%%%%%%%%%%%%%%%%%%%%%%%%%%%%%%%%%%%%%%%%%%%%%%%%%%%%%%%%%%%%%%%%%%%%%%%%%%%%%%%%%%%%%%%%%%

In this model, the qualitative behavior of both $\Omega$ and $\Omega_\Lambda$ is very similar to the previous fourth-order case, although the decrease in the value of the dark energy parameter occurs for later times and its minimum is in the near future instead of in the near past. We also verify that ${q{}}$ is monotonically decreasing in this case, starting from a large value in the far past and growing to a large negative value in the far future. This is again consistent with a late-time cosmic acceleration period, but in this case, instead of approaching a constant value, the acceleration is increasing. The behavior of ${q{}}$ also implies that this model cannot describe inflation, and that a dark energy dominated era is not directly related to accelerated rates of expansion. Notice that the behavior of $q$ implies that the model will accelerate faster and faster possibly leading to a big rip \cite{Caldwell:2003vq}.

%\newpage
%%%%%%%%%%%%%%%%%%%%%%%%%%%%%%%%%%%%%%%%%%%%%%%%%%%%%%%%%%%%%%%%%%%%%%%%%%%%%%%%%%%%%%%%%%%
\section{Static $H=0$ universes: Solutions at the infinite boundary of the phase space
and their stability}\label{Static}
%%%%%%%%%%%%%%%%%%%%%%%%%%%%%%%%%%%%%%%%%%%%%%%%%%%%%%%%%%%%%%%%%%%%%%%%%%%%%%%%%%%%%%%%%%

The variables we have defined in the previous section are efficient in determining the fixed points corresponding to a finite value of the quantities they represent. However, since variables have a term $H$ or $H^2$ in the denominator, our setting excludes an interesting case which is connected with the existence of solutions characterized by  $H=0$, i.e., static universes. In particular, fixed points, if any, associated to this kind of solution would be at the infinite boundary of the phase space. In order to look for solutions with $H=0$ one has, therefore, to investigate the asymptotics of the dynamical system. 

There are many approaches that can be adopted for this purpose. One could employ, for example, stereographic projections by which the infinite boundary is mapped to a finite radius sphere \cite{Perko}. In the following we will use a different strategy which allows us to analyze the stability of a static universe without having to explore the entire asymptotics. More specifically we will redefine all the variables and functions in such a way to bring the static fixed point into the finite part of the phase space. As said, the exploration of this extended phase space, is clearly not a complete analysis of the asymptotia, but it will allow an easier analysis of the stability of these solutions.

The cosmological parameters appearing in Eqs.~(\ref{qpar})-(\ref{spar}) are redefined as
\be\label{qbpar}
\overline q=\frac{H'}{\left(H+c\right)}\,,
\ee
\be\label{jbpar}
\overline j=\frac{H''}{\left(H+c\right)}\,,
\ee
\be\label{sbpar}
\overline s=\frac{H'''}{\left(H+c\right)}\,,
\ee
respectively,
where $c$ is an arbitrary constant with units of $H$.
The set of dynamical dimensionless variables in Eq.~(\ref{dynvar})
is also redefined as
\beq
&&\overline K=\frac{k}{a^2\left(H+c\right)^2},\ \ \ \overline X=\frac{\mathcal H}{\left(H+c\right)},\ \ \ \overline Y=\frac{R}{6\left(H+c\right)^2}, \ \ \ \overline Z=\frac{\mathcal R}{6\left(H+c\right)^2}, \nonumber \\ 
&&\ \overline Q=\overline q, \ \ \ \overline J=\overline j, \ \ \ \overline S=\overline s,\ \ \ \overline\Omega=\frac{\rho}{3\left(H+c\right)E},\ \ \ \overline A=\frac{R_0}{6\left(H+c\right)^2}, \ \ \ \overline T=\frac{H}{\left(H+c\right)}.\label{barvar}
\eeq
The Jacobian $J$ of this definition of variables can be written in the form,
\be\label{jacobian2}
J=\frac{1}{108a^2\left(H+c\right)^9E}\,.
\ee
This means that, for each specific model, the constraints that arise from imposing that the Jacobian must be finite and different from zero are the same as in the analysis of the previous section.
The evolution equations become in this case,
\beq
\overline K'&=&-2\overline K\left(\overline Q+\overline T\right)\,,\nonumber \\
\overline X'&=&\overline Z-\overline X\left(\overline Q+\overline X+\overline T\right)-
\overline K\,,\nonumber \\
\overline Y'&=&\overline J-2\overline K\overline T+\overline Q\left(4\overline T-2\overline Y+\overline Q\right)\,,\nonumber \\
\overline Z'&=&-\textbf{C}\left(\overline J+\overline Q^2-2\overline K\overline T+4\overline Q\overline T\right)+\textbf{D}\overline T^2\left(\overline X-\overline T\right)-2\overline Q\overline Z\,,\\
\overline Q'&=&\overline J-\overline Q^2\,, \nonumber \\
\overline J'&=&\overline S-\overline J\overline Q\,, \nonumber \\
\overline \Omega'&=&\frac{\overline \Omega}{\textbf{D}\overline T}\left\{-\textbf{D}\overline T^2\left[2\overline Q+3\overline T\left(1+w\right)\right]+2\textbf{A}\left[\left(\textbf{C}^2-\textbf{I}\right)\left(\overline J+\overline Q^2-2\overline K\overline T+4\overline Q\overline T\right)+\textbf{C}\textbf{D}\overline T^2\left(\overline T-\overline X\right)\right]\right\}\,,  \nonumber
\eeq
and the Friedmann equation
\begin{eqnarray}
\textbf{A}\left[2\left(\textbf{I}-\textbf{C}^2\right)\left(\overline J+\overline Q^2-2\overline K\overline T-4\overline Q\overline T\right)+2\textbf{D}\textbf{C}\overline T^2\left(\overline X-\overline T\right)+\textbf{D}T\left(\overline K+\overline X^2-\overline Z\right)\right]+&\nonumber\\
+\textbf{D}\overline T\left[\overline K+\left(1+\textbf{B}\right)\overline T^2-\overline Y-\overline \Omega\right]=&0\,.
\end{eqnarray}
The Raychaudhuri equation is too long to be reported here but can be computed easily. 

Using this new system of equations we can investigate the extended phase space. As we will see, we will be able to  find  all the previously discovered fixed points as points with $\overline T=1$ plus extra sets of fixed points with $H=0$ which will have $\overline T=0$. As an example of how to apply the new dynamical system approach, we will perform the analysis for the models given in 
Eqs.~\eqref{act1},~\eqref{act2}, and~\eqref{act3}, with the model given in~\eqref{act4}
being so long that we skip its presentation. In all the three cases, there is only one fixed point with $H=0$, which corresponds to the origin, i.e.,
\be
\mathcal O =\{\overline K=0,\overline X=0,\overline Y=0,\overline Z=0,\overline Q=0,\overline J=0,\overline \Omega=0,\overline T=0\}.
\ee
This result is somewhat expected. The fact that we are looking for fixed points with $H=0$ implies directly that $\overline T=0$. Then, both $\overline K=0$ and $\overline \Omega=0$ are invariant submanifolds. Now, if $H=0$, then $a$ is a constant and $\dot a=\ddot a=0$, from which $\overline Q=J\overline =0$. Since $k=0$ from $\overline K=0$, then we also have $R=0$ and therefore $\overline Y=0$. This been said, the only values of both $\overline Z$ and $\overline X$ that make $\overline X'=0$ and $\overline Z'=0$ using the results explained in this paragraph are $\overline Z=0$ and $\overline X=0$, and the fixed point is the origin. We now briefly comment on each model.

%%%%%%%%%%%%%%%%%%%%%%%%%%%%%%%%%%%%%%%%%%%%%%%%%%%%%%%%%%%%%%%%%%%%%%%%%%%%%%%%%%%%%%%%%%%
\paragraph*{Model $R^n\mathcal R^m$:}\hskip -0.2cm
For the model from Eq.~\eqref{act1} the fixed point $\mathcal O$ is always unstable, but might correspond to a saddle point or to a repeller depending on the parameters $n$ and $m$. In fact, if $1-m<n<0$ or $-m>n>0$, this point corresponds to a repeller. Any other combination of the parameters gives rise to a saddle point.

%%%%%%%%%%%%%%%%%%%%%%%%%%%%%%%%%%%%%%%%%%%%%%%%%%%%%%%%%%%%%%%%%%%%%%%%%%%%%%%%%%%%%%%%%%%
\paragraph*{Model $R^n+\mathcal R^m$:}\hskip -0.2cm
For the model from Eq.~\eqref{act2}, note that we have one extra variable $\overline A$ which also vanishes in this calculation. This implies that not all values of $m$ and $n$ are allowed, since the power of $\overline A$ must be positive for the system to converge. Despite that, the analysis of the stability in this case reveals that, for all the combinations of the parameters $m$, $n$ and $\gamma_*$ for which the fixed point exists, it is always a saddle point.

%%%%%%%%%%%%%%%%%%%%%%%%%%%%%%%%%%%%%%%%%%%%%%%%%%%%%%%%%%%%%%%%%%%%%%%%%%%%%%%%%%%%%%%%%%%%
\paragraph*{Model $\exp\left(\frac{R}{\mathcal R}\right)$:}\hskip -0.2cm
For the model from Eq.~\eqref{act3}, the analysis of the stability of the fixed point $\mathcal O$ reveals that it is unstable, since the eigenvalues associated to it are all either positive or zero. However, to verify if the fixed point is a saddle point or a repeller, one would have to make use of the central manifold theorem again. For our purposes however, it is enough for us to note that, since all the other eigenvalues of the point are of alternate sign, we can conclude directly that the point is unstable.

%%%%%%%%%%%%%%%%%%%%%%%%%%%%%%%%%%%%%%%%%%%%%%%%%%%%%%%%%%%%%%%%%%%%%%%%%%%%%%%%%%%%%%%%%%%%
\paragraph*{Model $R\exp\left(\frac{R}{\mathcal R}\right)$:}\hskip -0.2cm
For the model from Eq.~\eqref{act4} there appear to be no fixed points corresponding to an Einstein static universe.

%%%%%%%%%%%%%%%%%%%%%%%%%%%%%%%%%%%%%%%%%%%%%%%%%%
\section{Conclusions}\label{conc}
%%%%%%%%%%%%%%%%%%%%%%%%%%%%%%%%%%%%%%%%%%%%%%%%%

In this work we have applied the methods of dynamical systems to analyze the structure of the phase space of the generalized hybrid metric-Palatini gravity in a cosmological frame. Using the symmetries in the curvatures $R$ and $\mathcal R$ we obtained the cosmological equations of the theory. Then, defining the appropriate dynamical variables and functions, we derived a closed system of dynamical equations that allows us to study the phase space of different forms of the function $f\left(R,\mathcal R\right)$.  We studied four different models of the function $f$, namely, the ones given in Eqs.~\eqref{act1},~\eqref{act2},~\eqref{act3}, and~\eqref{act4}.

Independently of the model, the solutions for the scale factor can only be of two different kinds, depending on the cosmological snap parameter $S$ being zero or nonzero. We have shown that, if we assume that the Universe presents a vanishing snap parameter,
i.e., $S=0$, then the solution for the scale factor is analytical and presents a set of three integration constants that can be fine-tuned to yield the observed results for the Hubble parameter, the deceleration parameter, and the time interval since the big bang. This solution can also qualitatively model the inflation and the late-time cosmic acceleration periods. Furthermore, this solution also provides a prediction for the cosmological jerk parameter of ${j{}} \sim 4.47$. This prediction is of the same order of magnitude as the constraints imposed by the data of the Hubble parameter \cite{Mamon:2018dxf}. We should however emphasize that, as we have assumed that $S=0$, this solution corresponds to an approximation of the real solution which, in general, will present a non-anishing cosmological snap parameter.

The structure of the phase space is similar in all the four studied cases. In none of the particular cases there were global attractors due to the fact that one of the invariant submanifolds present in all the cases, $Z=0$, corresponds to a singularity in the phase space, and therefore a global attractor which would have to be in the intersection of all the invariant submanifolds would correspond to a singular state of the theory. On the other hand, the presence of these submanifolds allows us to discriminate sets of initial conditions and predict the time asymptotic states of the theory.

A fixed point that we denoted by $\mathcal B$ features in the theories given in Eqs.~\eqref{act1},~\eqref{act2}, and~\eqref{act4}, but not in the case~\eqref{act3}. This fixed point stands in the intersection of two of the invariant submanifold, $\Omega=0$ and $K=0$, with a positive value $Z=2$.
In the model from Eq.~\eqref{act1}  $\mathcal B$ is an attractor for some particular values of $n$ and $m$, see Table~\ref{fixed1}.
This means that all the orbits starting with a positive value of $Z$ and  with a value of $Y$ with the same sign as the one arising from the particular choice of parameters in $\mathcal B$, can eventually reach this fixed point. Note that it is even possible to chose sets of parameters such that $\mathcal B$ is the only finite attractor for the system, see Table~\ref{specific2}. The solution associated to point $\mathcal B$ is characterized by $s=0$ and contains three constants of integration $H_0$, $H_1$, and $H_2$. Depending of the values of these constants it can have two different types of asymptotic limit, namely,  a constant or a finite type singularity. The value of the constants  $H_0$, $H_1$, and $H_2$, and therefore the possibility of the occurrence of the singularity depends on observational constraints on higher order cosmological parameters (e.g., jerk, snap, and so on). This suggests that models for which $\mathcal B$ is an attractor in generalized hybrid-metric Palatini theories, like in many $f(R)$-gravity models, can incur in finite time singularities.
In the model of Eq.~\eqref{act2} $\mathcal B$ is always unstable. Note also that for this case given in Eq.~\eqref{act2}, no finite nor asymptotic attractors were found. One explanation for this result is that the orbits in the phase space do not tend asymptotically to a given solution but could instead be closed upon themselves. A structure similar to this one occurs for example in the frictionless pendulum, where the orbits are closed and there are no attractors in the phase space. This structure could indicate that the solutions represented by orbits in the phase space actually correspond to cyclic universes.
In the model of Eq.~\eqref{act4} the system simplifies further, see below for comments.

The fixed points $\mathcal E_\pm$ are
the other possible attractors in the phase space of the theories we have analyzed. They appear in the case of Eqs.~\eqref{act1} and~\eqref{act3}.
The fixed points $\mathcal E_\pm$ have always a solution $S\neq 0$ which asymptotically tends to a constant scale factor, a feature
that is absent for
the fixed point $\mathcal B$.
Moreover,
the fixed points $\mathcal E_\pm$ also lie in the intersection of the $\Omega=0$ and $K=0$ invariant submanifolds; thus we expect that some of the orbits should reach this fixed point. For the model in Eq.~\eqref{act1} we have shown that it is possible to choose sets of parameters such that $\mathcal E_+$ is the only finite attractor for the system, see Table~\ref{specific1}. On the other hand, for the model in Eq.~\eqref{act3} $\mathcal E_-$ is always the only finite attractor of the system, and all the orbits starting from a positive value of $Y$ and a negative value of $Z$ might reach this fixed point.

For the specific case shown in Eq.~\eqref{act4}, the system of dynamical equations becomes much simpler since only three variables are needed to fully describe the phase space and the solutions. However, the study of the stability becomes more complicated because the fixed points are not hyperbolic and their stability analysis requires the use of central manifolds. The behavior of the solutions reduces to simple power-laws or exponentials.
Our analysis connects directly with the paper \cite{Rosa:2017jld} in which some pairs function-exact solution were found via a reconstruction method. One of the limitation of the reconstruction technique was the impossibility to understand the stability of the solution obtained. The phase space analysis gives us a tool to determine this stability. In particular, we determined that the solutions found in \cite{Rosa:2017jld} are actually unstable.

We have also performed a numerical integration of the dynamical equations, starting from an initial condition consistent with the observations of the cosmological parameters, for both a fourth-order particular case of Eq.~\eqref{act1} and the second-order case of Eq.~\eqref{act4}, motivated by Ref.~\cite{Rosa:2017jld}. In the fourth order case we verified that, since there are no observational constraints on the cosmological snap parameter $S$, a degeneracy between the variables $Z$ and $X$ arises. In the second-order model, the cosmological parameter $S$ does not appear in the dynamical system and thus there are no degeneracies between dynamical variables in the initial state. Both the numerical integrations performed show that the dark energy contribution to the energy density is dominant both in the far past and into the future. However, there is a period for which the dark energy contribution decreases and the matter contribution increases. In the fourth-order model, this happens in the near past, whereas in the second-order model it happens in the near future. We have also shown that a dark-energy domination is not directly related to an accelerated expansion, as we have obtained that in the far past the deceleration parameter is positive and in the far future the deceleration parameter is negative, although both phases are associated to dark energy domination eras. Furthermore, we have verified that for our set of initial conditions the second-order model leads to a solution for which the deceleration parameter is negative and monotonically decreasing in the future, possibly leading to a big rip scenario.

The static $H=0$ universes were studied separately since
the variables in the set of equations given in Eq.~\eqref{dynvar} have $H$ in the denominator. Indeed,  the static $H=0$ fixed points are located at the asymptotic boundary of the phase space. In order to study these static universe solutions, we generalized Eq.~\eqref{dynvar} in such a way to move possible static fixed points  to the finite part of the phase space. This is different from a complete asymptotic analysis, but it allows us to obtain information on static universes in an easier way. All the theories we have considered  with Eq.~\eqref{dynvar} turn out to present a static fixed point which is always unstable. Therefore, as in general relativity, in these models the static universe is always unstable. However, differently from general relativity, the solution associated to these points is spatially flat and empty, i.e., with no cosmological constant. Such a peculiar form of static universes  is the result of the action of the nontrivial geometrical terms appearing in the field equations.
The existence of unstable static solutions in the phase space points to the existence in the context of generalized hybrid metric-Palatini gravity to phenomena such as bounces, turning points, and loitering phases, which are represented by the orbits bouncing against the static fixed points. These open the way to a series of scenarios which could be interesting to investigate further.

%%%%%%%%%%%%%%%%%%%%%%%%%%%%%%%%%%%%%%%%%%%%%%%%%%
\section*{Acknowledgements}
%%%%%%%%%%%%%%%%%%%%%%%%%%%%%%%%%%%%%%%%%%%%%%%%%

JLR acknowledges financial support from Funda\c{c}\~{a}o para a
Ci\^{e}ncia e Tecnologia FCT - Portugal for an FCT-IDPASC Grant
No.~PD/BD/114072/2015. SC acknowledges financial support by FCT
through Project No.~IF/00250/2013 and was also partly funded through H2020
ERC Consolidator Grant ``Matter and strong-field gravity: New
frontiers in Einstein's theory'', Grant Agreement No.~MaGRaTh-64659.
JPSL~acknowledges FCT for financial support through
Project~No.~UIDB/00099/2020.

%%%%%%%%%%%%%%%%%%%%%%%%%%%%%%%%%%%%%%%%%%%%%%%%%%%%
%\begin{thebibliography}{999}
%\bibitem{paper}
%\end{thebibliography}
%%%%%%%%%%%%%%%%%%%%%%%%%%%%%%%%%%%%%%%%%%%%%%%%%%%%

\bibliography{dynsys30.bib}

\end{document}